\documentclass[a4paper,11pt]{article}
\pdfoutput=1 

\usepackage{jheppub} 
\usepackage[T1]{fontenc} 
\usepackage{amsmath}
\usepackage{empheq}
\def\sgn{\text{sgn}}

\newcommand{\tl}[1]{\tilde{#1}}
\newcommand{\detr}{\text{det}\,}
\newcommand{\mc}[1]{\mathcal{#1}}
\def\[{\left[}
\def\]{\right]}
\def\({\left(}
\def\){\right)}
\def\cC{{\cal C}}

\def\cF{{\cal F}}

\def\cS{{\cal S}}

\def\cV{{\cal V}}

\newcommand{\be}{\beta}

\def\sgn {\text{sgn}}

\renewcommand{\ln}{\log}

\def\Tr{\mathrm{Tr}}

\def\nn{\nonumber\\}

\def\sgn{\text{sgn}}

\def\Tr{\mathrm{Tr}}

\def\nn{\nonumber\\}

\def \be {\begin{equation}}
\def \ee {\end{equation}}
\def \bea {\begin{eqnarray}}
\def \eea {\end{eqnarray}}
\def \beal#1 {\begin{align}#1\end{align}}

\everymath{\displaystyle}

\newcommand{\td}{\mathcal{D}}
\newcommand{\mbb}[1]{\mathbb{#1}}

\preprint{TIFR/TH/18-26} \title{\boldmath Duality and an exact
  Landau-Ginzburg potential for quasi-bosonic Chern-Simons-Matter
  theories}

 \author[a,1]{Anshuman
  Dey,\note{anshuman@theory.tifr.res.in}} \author[a,2]{Indranil
  Halder,\note{indranil.halder@tifr.res.in} } \author[b,3]{Sachin
  Jain,\note{sachin.jain@iiserpune.ac.in}} \author[a,4]{Lavneet
  Janagal,\note{lavneet@theory.tifr.res.in}} \author[a,5]{Shiraz
  Minwalla,\note{minwalla@theory.tifr.res.in}}\author[a,6]{Naveen
  Prabhakar \note{naveensp@theory.tifr.res.in}}
\affiliation[a]{Department of Theoretical Physics, Tata Institute of Fundamental Research, Homi Bhabha Rd, Mumbai 400005, India}
\affiliation[b]{Indian Institute of Science Education and Research, Homi Bhabha Rd, Pashan, Pune 411 008, India}

\abstract{ It has been conjectured that Chern-Simons (CS) gauged
  `regular' bosons in the fundamental representation are `level-rank'
  dual to CS gauged critical fermions also in the fundamental
  representation.  Generic relevant deformations of these conformal
  field theories lead to one of two distinct massive phases. In
  previous work, the large $N$ thermal free energy for the bosonic
  theory in the unHiggsed phase has been demonstrated to match the
  corresponding fermionic results under duality.  In this note we
  evaluate the large $N$ thermal free energy of the bosonic theory in
  the Higgsed phase and demonstrate that our results, again, perfectly
  match the predictions of duality. Our computation is performed in a
  unitary gauge by integrating out the physical excitations of the
  theory - i.e. W bosons - at all orders in the 't Hooft coupling. Our
  results allow us to construct an exact quantum effective potential
  for ${\bar \phi} \phi$, the lightest gauge invariant scalar operator
  in the theory. In the zero temperature limit this exact
  Landau-Ginzburg potential is non-analytic at ${\bar \phi
    \phi}=0$.
  The extrema of this effective potential at positive
  ${\bar \phi}\phi$ solve the gap equations in the Higgsed phase while
  the extrema at negative ${\bar \phi} \phi$ solve the gap equations in
  the unHiggsed phase. Our effective potential is bounded from below
  only for a certain range of $x_6$ (the parameter that governs sextic
  interactions of $\phi$). This observation suggests that the regular
  boson theory has a stable vacuum only when $x_6$ lies in this
  range.}

\begin{document} 
\maketitle
\flushbottom

\section{Introduction}
There is now considerable evidence that a single fermionic field in
the fundamental of $U(N_F)$ minimally coupled to $U(N_F)$ Chern-Simons
gauge theory at level\footnote{In our conventions the level of a
  Chern-Simons theory coupled to fermions is defined to be the level
  of the low energy gauge group obtained after deforming the theory
  with a fermion mass of the same sign as the fermion level.} $k_F$ is
dual to vector $SU(N_B)$ Wilson-Fisher scalars minimally coupled to
$SU(N_B)$ Chern-Simons gauge theory at level $ k_B=-{\rm sgn}(k_F)N_F$
with $N_B=|k_F|$ \cite{Choudhury:2018iwf}-\cite{Cordova:2017vab}
\footnote{See the introduction to the recent paper
  \cite{Choudhury:2018iwf} for a more more detailed description of
  earlier work.}. This (almost standard by now) duality asserts that
the two so-called \emph{quasi-fermionic} CFTs i.e.~Chern-Simons gauged
`regular fermions' (RF) and `critical bosons' (CB) - are secretly the same
theory.

It has also been conjectured (see \cite{Minwalla:2015sca} and
references therein) that the `quasi-fermionic' duality of the previous
paragraph follows as the infrared limit of a duality between pairs of
fermionic and bosonic RG flows. The fermionic RG flows are obtained by
starting in the ultraviolet with the Chern-Simons gauged Gross-Neveu
or `critical fermion' (CF) theory and deforming this theory with
relevant operators fine tuned to ensure that the IR end point of the
RG flow is the RF theory. In a similar manner the conjecturally dual
bosonic flows are obtained by starting in the ultraviolet with the
gauged `regular boson' (RB) theory deformed with the fine tuning that
ensures that the RG flow ends in the CB theory.

The UV starting points of the flows described above define dual pairs
of conformal field theories.  These RB and CF theories - so-called
\emph{quasi-bosonic} theories - are conjectured to be dual to each
other \footnote{At leading order in large $N$ - the order to which we
  work in this paper - the RB and CF theories appear as a line of
  fixed points parametrized by the single parameter $x_6$, the
  coefficient of the $\phi^6$ coupling of the bosonic theory (see
  below for the dual statement in the fermionic theory). In other
  words the one parameter set of RB and CF theories (and flows
  originating therein) that we study in this paper are actually only
  physical at three particular values of the parameter $x_6$. See the
  very recent paper \cite{abcd} for a computation of the beta function
  for $x_6$ that establishes this point.}.  If valid, this conjecture
implies that the set of all RG flows that originate in the RB theory
are dual to the set of all RG flows that originate in the CF theory.
The duality of the pair of specially tuned RG flows of the last
paragraph- those that end in the IR in the quasi-fermionic conformal
field theories - is a special case of this general phenomenon.

Generic RG flows that originate at quasi-bosonic fixed points lead to
gapped phases, or more accurately, phases whose low energy behaviour
is governed by a topological field theory.  There are two inequivalent
topological phases. In the unHiggsed phase the bosonic
(resp.~fermionic) theory is governed at long distances by pure
$SU(N_B)_{k_B}$ (resp.~$U(N_F)_{k_F}$) topological field theory with
the two theories being level-rank dual to each other. In the Higgsed
phase the bosonic (resp.~fermionic) theory is governed in the IR by a
pure $SU(N_B-1)_{k_B}$ (resp.~$U(N_F)_{{\rm sgn}\, k_F(|k_F|-1)}$)
Chern-Simons theory with the two topological field theories once again
being level-rank dual to each other. \footnote{As first explained in
  \cite{Aharony:2012nh}, the reduction in rank of the bosonic
  Chern-Simons theory compared to the unHiggsed phase is a consequence
  of the Higgs mechanism in the bosonic field theory. The reduction in
  level of the fermionic Chern-Simons theory is a consequence of the
  switch in sign of the mass of the fermion - level of the pure
  Chern-Simons theory obtained by integrating out a negative mass
  fermion is one unit smaller than the level obtained by integrating
  out a positive mass fermion. }

The most compelling evidence for the scenarios spelt out above comes
from explicit results of direct all-orders calculations that have been
performed separately in the fermionic and bosonic theories in the
large $N$ limit. In particular, the thermal partition function of
deformed RF and CB theories have both been computed in the unHiggsed
phases to all orders in the 't Hooft coupling, and have been shown to
match exactly with each other for all relevant deformations that end
up in this phase \cite{Minwalla:2015sca} \footnote{A similar
  matching has also been performed for the S-matrix in the unHiggsed
  phase \cite{Jain:2014nza, Yokoyama:2016sbx}. The generalisation of
  this match to the Higgsed phase is also an interesting project, but
  one that we will not consider in this paper.}.  While impressive,
this matching is incomplete, as the restriction to the unHiggsed
phase covers only half of the phase diagram of these theories.

The authors of \cite{Minwalla:2015sca} (and references therein) were
also able to compute the thermal partition function of the CF theory
in the `Higgsed' phase. However, they were unable to perform the
analogous computation in the bosonic theory in this phase and so were
unable to verify the matching of thermal free energies in this phase. In this paper we fill the gap described above. We present an explicit
all-orders computation of the thermal free energy of the RB theory
in the Higgsed phase. Under duality our final results exactly match
the free energy of the fermionic theory in the Higgsed phase, completing
the large $N$ check of the conjectured duality in a satisfying manner.

At the technical level, the computation described in the previous 
paragraph (and presented in detail in section \ref{higgsphase}) is a relatively
straightforward generalisation of the computations presented in the
recent paper \cite{Choudhury:2018iwf}. In \cite{Choudhury:2018iwf} the
large $N$ free energy of the Higgsed phase of the Chern-Simons gauged
Wilson-Fisher boson theory was computed for the first time.  As we
describe in much more detail below, the computation of the free energy
in the Higgsed phase of the RB theory can be divided into two
steps. In the first step we compute the thermal free energy (or
equivalently, the gap equation) of the CB theory as a function of its
Higgs vev. We are able to import this computation directly from
\cite{Choudhury:2018iwf}.  In the relatively simple second step
carried through in this paper, we derive a second gap equation
that determines the effective value of the Higgs vev.

The second step described at the end of the last paragraph had no
counterpart in \cite{Choudhury:2018iwf}. In the critical bosonic
theory the `classical' potential for the scalar field is infinitely
deep. This potential freezes the magnitude of the scalar field in the
Higgsed phase to its classical minimum even in the quantum theory. It
follows that the Higgs vev is independent of the temperature and has a
simple dependence on the 't Hooft coupling in the critical boson
theory. In the regular boson theory, on the other hand, the classical
potential for the scalar field is finite and receives nontrivial
quantum corrections. The value of the scalar condensate is determined
extremizing the quantum effective action for the scalar field. The
result of this minimisation yields a scalar vev that is a nontrivial
function of both the 't Hooft coupling and the temperature.  It
follows that the computations of this paper give us a bonus: we are
able to compute the smooth `quantum effective potential' for the RB
theory as a function of the Higgs vev. More precisely we compute the
quantum effective potential for the composite field
$({\bar \phi \phi})$. In the Higgsed phase and in the unitary gauge
employed in the computations of this paper, this quantity reduces to a
potential - an exact Landau-Ginzburg effective potential - for the
Higgs vev. The extremization of this potential determines the Higgs
vev - and is an equivalent and intuitively satisfying way of obtaining
the gap equations - in the Higgsed phase.  The later sections of this
paper - sections \ref{ose} and \ref{ost} are devoted to the study of
the exact quantum effective potential of the theory and its physical
consequences.

Let us denote the expectation value of $({\bar \phi \phi})$ by
$({\bar \phi \phi})_{\rm cl}$. At zero temperature it turns out that
the quantum effective action is non-analytic at
$({\bar \phi \phi})_{\rm cl}=0$. For this reason the domain of the
variable in our effective potential - namely ${\bar \phi \phi}$ -
naturally splits into two regions. We refer to the region
${\bar \phi \phi} >0$ as the Higgsed branch of our effective
potential. On the other hand the region ${\bar \phi \phi} <0$ is the
unHiggsed branch of our effective potential. On the Higgsed branch the
quantum effective potential for ${\bar \phi \phi}$ is simply a quantum
corrected version of the classical potential of the
theory. Classically ${\bar \phi \phi}$ always positive, and so the
potential for the theory on the unHiggsed branch (i.e.~at negative
${\bar \phi \phi}$) has no simple classical limit and is purely
quantum in nature.  The extremization of the effective potential on
the Higgsed/unHiggsed branches exactly reproduces the gap equations in
the Higgsed/unHiggsed phases.

In both phases the extrema of this effective potential are of two
sorts; local maxima and local minima. Local maxima clearly describe
unstable `phases'. The instability of these phases has an obvious
semiclassical explanation in the Higgsed phase; it is a consequence of
the fact that we have chosen to expand about a maximum of the
potential for the Higgs vev. In this paper we find an analogous
physical explanation for the instability of the `maxima' in the
unHiggsed phase.  In Section \ref{uhp} we use the results for exact
S-matrices in these theories \cite{Jain:2014nza, Yokoyama:2016sbx} to
demonstrate that the `phases' constructed about maxima in the
unHiggsed branch always have bound states of one fundamental particle
(created by $\phi$) and one antifundamental particle (created by
${\bar \phi}$) in the so called `singlet' channel. Moreover we
demonstrate in Section \ref{uhp} that these bound states are always
tachyonic (i.e. have negative squared mass).  As a consequence, such
expansion points are maxima in the potential of the field that creates
these bound states (in this case ${\bar \phi } \phi$), explaining the
instability of the corresponding solutions of the gap equation.

It turns out that our effective potential is unbounded from below in
the limit $({\bar \phi \phi})_{\rm cl} \to +\infty$ when
$x_6< \phi_1$. Here $x_6$ is the parameter that governs the $\phi^6$
interaction of the theory defined precisely in \eqref{rst}, and
$\phi_1$ is a particular function of the 't Hooft coupling $\lambda_B$
of this theory listed in \eqref{phi12def}. When
$({\bar \phi \phi})_{\rm cl} \to -\infty$, on the other hand, the
potential turns out to be unbounded from below when $x_6> \phi_2$;
$\phi_2$ is given in \eqref{phi12def}. It follows that the RB theory
is unstable - i.e.  does not have a stable vacuum state - if either of
the conditions above are met. Happily it turns out that
$\phi_1<\phi_2$ so that there is a range of values for $x_6$, namely
\begin{equation}
\label{stabrange}
 \phi_1 \leq x_6 \leq \phi_2\ ,
 \end{equation}
over which the regular boson theory is stable. 

The zero temperature phase diagram of the RB theory was worked out in
great detail in the recent paper \cite{abcd}. In order to accomplish
this, the authors of \cite{abcd} evaluated every solution of the gap
equation of the RB theory and then compared their free energies. The
dominant phase at any given values of microscopic parameters is simply
the solution with the lowest free energy; this dominant solution was
determined in \cite{abcd} by performing detailed computations.  In
Section \ref{phase} of this paper we demonstrate that the structure
of the phase diagrams presented in \cite{abcd} has a simple intuitive
explanation in terms of the exact Landau-Ginzburg effective potential
for $({\bar \phi} \phi)_{\rm cl}$ described above. As we explain in
Section \ref{phase} below, the general structure of the phase diagram
follows from qualitative curve plotting considerations and can be
deduced without performing any detailed computations. Moreover the
analysis of the current paper has an added advantage; it allows us to
distinguish regions of the phase diagram where the dominant phase is
merely metastable (this happens when $x_6>\phi_2$ or $x_6<\phi_1$)
from regions in the phase diagram in which the dominant solution of
the gap equation is truly stable (this happens in the range
\eqref{stabrange}).

\section{Review of known results} 
\subsection{Theories and the conjectured duality map}
The RB theory is defined by the action
\begin{align}
	S_B  &= \int d^3 x  \biggl[i \varepsilon^{\mu\nu\rho}{\kappa_B\over 4\pi}
	\Tr( X_\mu\partial_\nu X_\rho -{2 i\over3}  X_\mu X_\nu X_\rho)
	 + D_\mu \bar \phi D^\mu\phi  \nn
	&\qquad\qquad\quad
	+m_B^2 \bar\phi \phi +  {4\pi { b}_4 \over \kappa_B}(\bar\phi \phi)^2
	+ \frac{(2\pi)^2}{\kappa_B^2} \left( x_6^B+1\right)  (\bar\phi \phi)^3\biggl],
	\label{rst}
\end{align}
while the $\zeta_F$ and $\zeta_F^2$ deformed critical fermion (CF)
theory is defined by the Lagrangian
\begin{align}\label{csfnonlinear}
  S_F  &= \int d^3 x \bigg[ i \varepsilon^{\mu\nu\rho} {\kappa_F \over 4
         \pi} \Tr( X_\mu\partial_\nu X_\rho -{2 i\over3} X_\mu X_\nu
         X_\rho) + \bar{\psi} \gamma_\mu D^{\mu} \psi \nn 
  &\qquad\qquad\quad
    -\frac{4\pi}{\kappa_F}\zeta_F (\bar\psi \psi - {\kappa_F
    y_2^2\over4\pi} ) - { 4\pi y_4\over \kappa_F} \zeta_F^2 +
    \frac{(2\pi)^2}{\kappa_F^2} x_6^F \zeta_F^3\bigg]\ .
\end{align}
In these formulae
\begin{equation}\label{kkmt}
\kappa_B={\sgn}(k_B) \left( |k_B|+N_B \right), \quad\kappa_F
= {\sgn}(k_F) \left( |k_F|+N_F \right)\ .
\end{equation}
The levels $k_F$ and $k_B$ are defined to be the levels of the WZW
theory dual to the pure Chern-Simons theory (throughout this paper we
work with the dimensional regularisation scheme). For concreteness, in
this paper we will assume that the bosonic theory gauge group
is $SU(N_B)$ while the fermionic gauge group is $U(N_F)$ with `equal'
levels $k_F$ for the $SU(N_F)$ and $U(1)$ parts of the gauge
group. The generalisation to $U(N_B) \leftrightarrow SU(N_F)$ and
$U(N_B) \leftrightarrow U(N_F)$ dualities is straightforward at large
$N$ and will not be explicitly considered in this paper.

In the
rest of this paper we will present our formulae in terms of the 't
Hooft couplings defined by
\begin{equation}\label{tcdef}
\lambda_B= \frac{N_B}{\kappa_B},~~~ \lambda_F= \frac{N_F}{\kappa_F}\ .
\end{equation}
We have already mentioned in the introduction that the two theories
above have been conjectured to be dual to each other under the
level-rank duality map
\begin{equation} \label{lgdm}
N_B=|\kappa_F|-N_F,~~~
\kappa_B= -\kappa_F\ .
\end{equation}  
This implies that the bosonic 't Hooft coupling is given in terms of
its fermionic counterpart by
\begin{equation}\label{tcm}
\lambda_B=\lambda_F-{\rm sgn}(\lambda_F), \quad
\end{equation}
The relations \eqref{lgdm} and \eqref{tcm} are expected to hold even
at finite $N$. On the other hand the map between deformations of these
two theories is conjectured to be\footnote{Since $x_6^B = x_6^F$, we
  drop the superscript ${}^B$ or ${}^F$ on $x_6$ often in the paper
  when referring to this coupling.}
\begin{equation}
  x_6^F =x_6^B\ , \quad y_4 = { b}_4\ , \quad
  y_2^2 = m_B^2\ .
\label{dualitytransform}
\end{equation}
The above equation \eqref{dualitytransform} is known to hold only in
the large $N$ limit; this relationship may well receive corrections in
a power series expansion in $\frac{1}{N}$.

To end this subsection, let us note that under the field redefinition 
\begin{equation}\label{psidef}
\phi= \sqrt{\kappa_B}\, \varphi\ ,
\end{equation}
the action \eqref{rst} turns into
\begin{align}
S_B &=  \frac{N_B}{\lambda_B} \int d^3 x  \biggl[i \varepsilon^{\mu\nu\rho}{1\over 4\pi}
\Tr( X_\mu\partial_\nu X_\rho -{2 i\over3}  X_\mu X_\nu X_\rho) + D_\mu \bar \varphi D^\mu\varphi \nonumber\\ 
&\qquad\qquad\qquad\qquad + m_B^2 \bar\varphi \varphi +  4\pi { b}_4 (\bar\varphi \varphi)^2
+ (2\pi)^2 \left( x_6^B+1\right)  (\bar\varphi \varphi)^3\biggl]\ .
\label{rstr}
\end{align}
It follows immediately that in the  limit 
\begin{equation}\label{class}
\lambda_B \to 0\ ,\quad m_B^2,~b_4,~x_6 ={\rm fixed}\ ,
\end{equation}
the theory \eqref{rst} should reduce to a nonlinear but classical 
theory of the fields $\varphi$ and $X_\mu$. We will return to this 
point below.

\subsection{Structure of the thermal partition function}
As explained in e.g.~\cite{Choudhury:2018iwf}, the large $N$ thermal
free energy of either of these theories on $S^2 \times S^1$ can be
obtained following a two step process. In the first step we compute
the free energy of the theory in question on
${\mathbb R}^2 \times S^1$,  at a fixed value of the gauge
holonomies around $S^1$. The result is a functional of the holonomy
eigenvalue distribution function $\rho(\alpha)$ and is given by the
schematic equation
\begin{equation}\label{nseff}
e^{-\mc{V}_2 T^2 v[\rho] } = \int_{\mbb{R}^2 \times S^1} [d \phi]\ e^{-S[\phi, \rho]}\ .
\end{equation}
where $\mc{V}_2$ is the volume of two dimensional space and $T$ is the
temperature.

In order to complete the evaluation of the $S^2 \times S^1$ partition
function of interest, in the next step we are instructed to evaluate
the unitary matrix integral
\begin{equation}\label{nioev}
\mc{Z}_{S^2\times S^1} =\int [dU]_{\text{CS}} ~e^{-\mc{V}_2 T^2 v[\rho]}.
\end{equation}
where $[dU]_{\text{CS}}$ is the Chern-Simons modified Haar measure
over $U(N)$ described in \cite{Jain:2013py}. 

It was demonstrated in \cite{Jain:2013py} that the thermal partition
functions \eqref{nioev} of the bosonic and fermionic theories agree
with each other in the large $N$ limit provided that under duality
\begin{equation}\label{mv}
  v_F[\rho_F]= v_B[\rho_B]\ ,
\end{equation}
where the bosonic and fermionic eigenvalue distribution functions,
$\rho_B$ and $\rho_F$, are related via
\begin{equation}\label{nred}
|\lambda_B| \rho_B (\alpha)+ |\lambda_F| \rho_F(\pi - \alpha) = \frac{1}{2 \pi}.
\end{equation}
In this paper we will evaluate the free energy $v_B[\rho_B]$ of the
bosonic theory in the Higgsed phase and verify \eqref{mv}, thus
establishing the equality of thermal free energies of the RB and CF
theories in the Higgsed phase. We summarise the map between the
parameters \eqref{lgdm}, \eqref{tcm}, \eqref{dualitytransform} and the
holonomy distributions \eqref{nred}:
\begin{align}\label{dualitymapmain}
&N_F = |\kappa_B| - N_B\ ,\quad \kappa_F = - \kappa_B\ ,\quad \lambda_F = \lambda_B - \sgn(\lambda_B)\ ,\nonumber\\
&x_6^F = x_6^B\ ,\quad y_4 = b_4\ ,\quad y_2^2 = m_B^2\ ,\quad |\lambda_B| \rho_B (\alpha)+ |\lambda_F| \rho_F(\pi - \alpha) = \frac{1}{2 \pi}.
\end{align}

In Appendix \ref{review} we provide a comprehensive review of
everything that is known about the large $N$ thermal free energies of
the CF and RB theories (the appendix also contains a formula for a
`three variable off-shell' free energy functional of the CF theory
that is valid in both phases \eqref{ofcrf12}). In the rest of this
section we only present those results that will be of relevance for
the computations in the paper.

The free energy $v_F[\rho_F]$ in the critical fermion theory has been
computed in both fermionic phases in \cite{Minwalla:2015sca}. The
result is given in terms of an auxiliary off-shell free
energy\footnote{ We put a hat over a particular quantity
  (e.g.~${\hat c}_F$) to denote the dimensionless version of that
  quantity (e.g.~$c_F$) obtained by multiplying by appropriate powers
  of the temperature $T$.} (equation \eqref{ofcrf} in Appendix \ref{review})
\begin{align}\label{ofcrfintro}
  F_F(c_F, \zeta_F) &=\frac{N_F }{6\pi} \bigg[ \frac{|\lambda_F| - \sgn(\lambda_F)\sgn(X_F)}{|\lambda_F|} \hat{c}_F^3 - \frac{3}{2\lambda_F}\left(\frac{4\pi}{\kappa_F} \hat\zeta_F\right) \hat{c}_F^2 \nonumber \\ &\qquad\qquad  + \frac{1}{2\lambda_F}\left(\frac{4\pi}{\kappa_F} \hat\zeta_F\right)^3 + \frac{6\pi {\hat y}_2^2}{\kappa_F\lambda_F}{\hat \zeta}_F  -\frac{24 \pi^2 {\hat y}_4}{\kappa_F^2\lambda_F}{\hat \zeta}_F^2 +\frac{24 \pi^3 x^F_6}{\kappa_F^3\lambda_F}{\hat \zeta}_F^3\nonumber \\  &\qquad\qquad - 3 \int_{-\pi}^{\pi}d\alpha \rho_{F}(\alpha)\int_{{\hat c}_F}^{\infty} dy ~y~\left( \ln\left(1+e^{-y-i\alpha}\right)  + \ln\left(1+e^{-y+i\alpha}\right)\right)\bigg].
\end{align}
The above free energy is a function of two variables $c_F$ and
$\zeta_F$. Extremizing $F_F$ with respect to these variables and
plugging back in the extremum values gives us the free energy
$v_F[\rho_F]$. The physical interpretation of the variable $c_F$ is
that its value at the extremum of $F_F$ coincides with the pole mass
of the fermion.

The free energy \eqref{ofcrfintro} assumes two different analytic
expressions depending on the sign $\sgn(\lambda_F) \sgn(X_F)$ and
governs the dynamics of the two different phases. The phase in which
${\rm sgn}({ X}_F) \sgn(\lambda_F) = \pm 1$ is referred to as the
unHiggsed phase and the Higgsed phase respectively. In equation
\eqref{ofcrf12} in Appendix \ref{review}, we give an off-shell free energy in terms
of three variables (which include $c_F$ and $\zeta_F$) which is
analytic in all three variables and encompasses the behaviour of both
phases.

The free energy \eqref{ofcrfintro} in the unHiggsed phase of the CF
theory matches the free energy of the regular boson theory in the
unHiggsed phase (equation \eqref{noffb12} in Appendix \ref{review}) computed in
\cite{Minwalla:2015sca} under the duality map
\eqref{dualitymapmain}. The free energy \eqref{ofcrfintro} with
$\sgn(\lambda_F)\sgn(X_F) = -1$ gives a prediction for the regular
boson theory in the Higgsed phase. Applying the duality transformation
\eqref{dualitymapmain} and making the following `field' redefinitions:
\begin{equation}
c_F = c_B\ ,\quad \frac{4\pi \zeta_F}{\kappa_F} = -2\lambda_B\sigma_B\ ,
\end{equation}
we get the following prediction for the free energy in the Higgsed
phase (equation \eqref{ferdualm}):
\begin{align}\label{ferdualmintro}
 F_B(c_B,\sigma_B) & = \frac{N_B }{6\pi} \bigg[- \frac{\lambda_B - 2\sgn(\lambda_B)}{\lambda_B} {\hat c}_B^3  - 3 {\hat \sigma}_B (\hat{c}_B^2 - \hat{m}_B^2) + 6{\hat b}_4 \lambda_B{\hat\sigma}_B^2  + (3  x^B_6 + 4)\lambda_B^2{\hat\sigma}_B^3\nonumber \\  &\qquad\qquad + 3 \int_{-\pi}^{\pi}d\alpha \rho_{B}(\alpha)\int_{{\hat c}_B}^{\infty} dy ~y\left( \ln\left(1-e^{-y-i\alpha}\right)  + \ln\left(1-e^{-y+i\alpha}\right)\right)\bigg].
\end{align}
The extremum value of the variable $c_B$ corresponds to the pole mass
of the W boson excitation in the Higgsed phase. In the next section we
will independently compute the off-shell free energy of the RB theory,
and demonstrate that our answer agrees with \eqref{ferdualmintro} once
we identify the field $\sigma_B$ with
\begin{equation}\label{mdefnintro}
 \sigma_B = 2\pi \frac{ {\bar \phi \phi}}{N_B}\ .
\end{equation}
where ${\bar \phi}$ and $\phi$ respectively stand for the saddle point
values of the corresponding fields denoted by the same letters (recall
these fields have nonzero saddle point values in the Higgsed phase).

\textbf{Note:} Here and in the rest of the paper, we define the
quantities $c_F$ and $c_B$ to be always positive. In other words,
$c_{F,B}$ is shorthand for $|c_{F,B}|$. This is the same
convention used in \cite{Minwalla:2015sca}.

\section{The Higgsed Phase of the regular boson theory}
\label{higgsphase}

\subsection{Lagrangian in Unitary gauge}
Consider the following action for the $SU(N_B)$ regular boson theory:
\begin{equation} \label{regboselag}
	\begin{split}
          S_{\text{E}}& =\int d^3x  \Big[\ i \epsilon^{\mu \nu \rho} \frac{\kappa_B}{4 \pi}\, \Tr(X_\mu \partial_\nu X_\rho-\frac{2 i}{3} X_\mu X_\nu X_\rho) + D_\mu\bar{\phi}D^\mu \phi\\
           &\qquad\qquad\quad +  m_B^2 \bar\phi \phi+\frac{4\pi b_4}{\kappa_B}(\bar\phi
          \phi)^2+\frac{(2\pi)^2}{\kappa_B^2}(x_6^B + 1)(\bar\phi
          \phi)^3\Big]\ ,
	\end{split}
\end{equation}
with $D_\mu = \partial_\mu - i X_\mu$. The above action can be
reorganised as follows in the Higgsed phase where we anticipate
$\langle \bar\phi \phi \rangle \neq 0$. Following
\cite{Choudhury:2018iwf} we work in the unitary gauge
\begin{equation}\label{unitary}
\phi^i(x) = \delta^{iN_B} \sqrt{|\kappa_B|}\, V(x)\ .
\end{equation}
For future reference we note also that \eqref{unitary} implies the following for the `classical' field $\varphi$ defined in \eqref{psidef}:
\begin{equation}\label{unitaryc}
\varphi^i(x) = \delta^{iN_B} \, \sqrt{{\rm sgn}(\kappa_B)}\, V(x)\ .
\end{equation}

The field $V(x)$ shall be termed the \emph{Higgs field}. The above
gauge choice lets us decompose the gauge field $X_\mu$ as
\begin{equation}
X_\mu = \begin{pmatrix} (A_\mu)^a{}_b - \tfrac{ \delta^a{}_b}{N_B-1} Z_\mu & \tfrac{1}{\sqrt{\kappa_B}}(W_\mu)^a \\ \tfrac{1}{\sqrt{\kappa_B}}(\bar{W}_\mu)_b & Z_\mu \end{pmatrix}\ ,
\end{equation}
where the indices $a, b$ run over $1,\ldots, N_B-1$. In terms of these
variables, the action can be rewritten as follows\footnote{The
  notation $ABC$ stands for
  $d^3 x \epsilon^{\mu\nu \rho} A_\mu B_\nu C_\rho$.}:
\begin{align}\label{asb}
S_{\text{E}}[A,W,Z,V]&=\frac{i \kappa_B}{4 \pi}\int \Tr(AdA-\tfrac{2i}{3}AAA) \nonumber\\ 
&\quad +\frac{i }{4 \pi}\int\left(2 \bar{W}_{a}d  W^a + \kappa_B Zd Z - 2iZ\bar{W}_a W^a - 2 i \bar{W}_aA^a{}_b W^b\right) \nonumber\\ 
&\quad +\int d^3x\, (|\kappa_B| V^2 Z_\mu Z^\mu + {\sgn}(\kappa_B) V^2\bar{W}_{a\mu} W^{a\mu}) \nonumber\\
&\quad+|\kappa_B|\int d^3x\left( \partial_\mu V \partial^\mu V +   m_B^2 V^2  + 4\pi b_4\sgn(\kappa_B) V^4 + 4\pi^2  (x^B_6 + 1) V^6\right)\ .
\end{align}

\subsection{An effective action for the Higgs field $V$}
We will now compute the thermal partition function of the 
regular boson theory in the Higgsed phase, i.e.~we will compute 
$v_B[\rho_B]$ defined by
\begin{equation}\label{nseffrb}
e^{-\mc{V}_2 T^2 v_B[\rho_B] } = \int_{\mbb{R}^2 \times S^1} 
[dV dW dZ dA]\ e^{-S_{\text{E}}[A,W,Z,V]}\ ,
\end{equation}
where $S_{\text{E}}[A,W,Z,V]$ was defined in \eqref{asb}. For this
purpose it is convenient to break up the effective action
$S_{\text{E}}[A,W,Z,V]$ into two parts
\begin{equation}
\label{betp}
S_{\rm E}[A,W,Z,V]= S_1[A,W,Z,V] + S_2[V]
\end{equation}
where 
\begin{align}\label{asb1}
S_1[A,W,Z,V]&=\frac{i \kappa_B}{4 \pi}\int d^3x \ \Tr(AdA-\tfrac{2i}{3}AAA) \nonumber\\ 
&\quad +\frac{i }{4 \pi}\int \left(2 \bar{W}_a d W^a + \kappa_B ZdZ - 2iZ\bar{W}_a W^a-2 i \bar{W}_aA^a{}_b W^b\right) \nonumber\\ 
&\quad +\int d^3x\, (|\kappa_B| V^2 Z_\mu Z^\mu + {\sgn}(\kappa_B) V^2\bar{W}^\mu_{a} W^{a}_\mu)\ ,
\end{align}
and 
\begin{align}\label{vpot}
  S_2[V] &= \int d^3x \Big(|\kappa_B| \partial_\mu V \partial^\mu V +  U_{\rm cl}(V)\Big)\ ,\nonumber\\
  U_{\rm cl}(V) &= |\kappa_B| m_B^2 V^2  + 4\pi b_4\kappa_B V^4 + 4\pi^2 |\kappa_B| (x^B_6 + 1) V^6\ .
\end{align}
The path integral \eqref{nseff} can be rewritten as 
\begin{equation}\label{nseffn}
e^{-\mc{V}_2 T^2 v_B[\rho_B] } = 
\int [dV]\, e^{-S_2[V]} \int [dW dZ dA]\, e^{-S_1[A,W,Z,V]}   
\end{equation}
Let us first study `inner' path integral i.e. 
\begin{equation}
\label{pie}
e^{-\mc{V}_2 T^2 v_{\rm CB}[\rho_B, V]} \equiv \int [dW dZ dA]\, e^{-S_1[A,W,Z,V]}\ , 
\end{equation} 
where the right hand side defines the quantity
$v_{\rm CB}[\rho_B, V]$. As far as the path integral in \eqref{pie} is
concerned, $V(x)$ is a background field. The path integral \eqref{pie}
is difficult to evaluate for arbitrary $V(x)$ even in the large $N_B$
limit\footnote{\eqref{pie} is effectively the generating function of
  all correlation functions of the dimension two scalar $J_0$ in the
  large $N$ critical boson theory, and so contains a great deal of
  information.}. This problem simplifies, however, in the special case
that $V(x)$ is a constant. In fact, precisely in this limit, the path
integral \eqref{pie} has been evaluated in the recent paper
\cite{Choudhury:2018iwf}. Luckily, it will turn out that, in the large
$N$ limit, the integral over $V(x)$ in \eqref{nseffn} localises to a
saddle point at which $V(x)$ is constant (see below). As a consequence
we only need the result of the path integral \eqref{pie} for constant
$V(x)$; we are able to read off this result directly from
\cite{Choudhury:2018iwf} which we now pause to recall.

The authors of \cite{Choudhury:2018iwf} studied the critical boson
theory in its Higgsed phase. Working in unitary gauge and following
manipulations essentially identical to those outlined in the previous
subsection, they found that the CB theory in the Higgsed phase can be
rewritten as effective theory of interacting massive $W$ bosons, $Z$
bosons and $SU(N_B-1)$ gauge fields, whose action is given by
\begin{align}\label{oldact}
S_{\text{E}}[A,W,Z]&=\frac{i \kappa_B}{4 \pi}\int \Tr(AdA-\tfrac{2i}{3}AAA) \nonumber\\ 
&\quad +\frac{i }{4 \pi}\int\left(2 \bar{W}_a d W^a + \kappa_B ZdZ - 2iZ\bar{W}_a W^a-2 i \bar{W}_aA^a{}_b W^b\right) \nonumber\\ 
&\quad -\int d^3x\, \left(\frac{N_B}{4\pi} m_B^{\text{cri}} Z_\mu Z^\mu + \frac{\lambda_B}{4\pi} m_B^{\text{cri}} \bar{W}_{a\mu} W^{a\mu}\right)\ .
\end{align}
The authors of \cite{Choudhury:2018iwf} were then able to evaluate the
finite temperature partition function for the theory defined by
\eqref{oldact}. Their final result for $v_{\rm CB}[\rho_B]$ is given
as follows. One obtains $v_{\rm CB}[\rho_B]$ by extremizing an
off-shell free energy $F_{\rm CB}(c_B)$ given by
\begin{equation} \label{osfeeh}
\begin{split}
  F_{\rm CB}(c_B) &=\frac{N_B}{6\pi} {\Bigg[} -
  \frac{ \lambda_B-2{\rm sgn}(\lambda_B)  }{\lambda_B}\hat{c}_B^3 +\frac{3}{2} {\hat m}_B^{\rm cri} \hat{c}_B^2 + \Lambda\left({\hat m}_B^{\rm cri}\right)^3\\
  &+3 \int_{-\pi}^{\pi} d\alpha\rho_B(\alpha)
  \int_{\hat{c}_B}^{\infty} dy\,
  y\left(\ln\left(1-e^{-y-i\alpha}\right)+\ln\left(1-e^{-y+i\alpha}\right)
  \right)
  {\Bigg]},\\
\end{split}
\end{equation}
Here, $\Lambda$ is an undetermined constant; shifts in $\Lambda$
correspond to shifts in the cosmological constant counterterm in the
starting action for the CB theory (see \cite{Choudhury:2018iwf} for a
discussion).

Note that the action $S_1$ in
\eqref{asb} agrees precisely with the action \eqref{oldact} reported
in \cite{Choudhury:2018iwf} if we replace $m_B^{\rm cri}$ by the quantity
\begin{equation}\label{reprule}
  m_B^{\text{cri}} = -\frac{4\pi}{|\lambda_B|} V^2\quad\text{with $V$ constant}.
\end{equation}
It follows that for the special case that $V(x)$ is constant, the path
integral \eqref{pie} is given by the extremum value of \eqref{osfeeh}
with the replacement \eqref{reprule} and the path integral over $V$ in
\eqref{nseffn} takes the form
\begin{equation}
\label{piv}
\int [dV] e^{-S_{\rm eff} [V]}\quad\text{with}\quad  S_{\rm eff}[V] = S_2[V] + \cV_2 T^2 v_{\rm CB}[\rho_B, V]\ .
\end{equation}
From the expressions for $S_2[V]$,\footnote{In
  \cite{Choudhury:2018iwf}, the free energy $v_{{\rm CB}}[\rho_B]$
  depended on $m_B^{\rm cri}$ which was a parameter in the
  theory. After the replacement \eqref{reprule}, the dependence on the
  parameter $m_B^{\rm cri}$ is replaced by a dependence on the field
  $V(x)$. We have included an explicit $V$ in the notation for
  $v_{\rm CB}[\rho_B, V]$ to highlight this dependence on the Higgs
  field $V$. } and $F_B(c_B)$ in \eqref{vpot} and \eqref{osfeeh}, it
is clear that there is an overall factor of $N_B$ in front of the
effective action $S_{\rm eff}[V]$. In the large $N_B$ limit the path
integral over $V$ may be evaluated in the saddle-point
approximation. We expect the dominant minima of the effective action
to occur at constant values of $V$ since the kinetic term
$\partial_\mu V \partial^\mu V$ adds a positive definite piece to the
action. For this reason it is sufficient to have the expression
$v_{\rm CB}[\rho_B, V]$ only at constant $V$. As we have already
explained above, this result is given by extremizing \eqref{osfeeh}
w.r.t.~$c_B$ after making the replacement \eqref{reprule}. It follows
that the final result for $v_B[\rho_B]$ in \eqref{nseffrb} is obtained
by extremizing the regular boson off-shell free energy
$$F_B(c_B, V) = F_{\rm CB}(c_B) + \frac{1}{\cV_2 T^2}S_2[V]\ ,$$ with respect
to both $c_B$ ${\it and }$ $V$.\footnote{It is understood that
  $F_{\rm CB}[\rho_B, c_B]$ is evaluated after making the
  replacement \eqref{reprule}. } Using the explicit expressions
\eqref{vpot}, \eqref{osfeeh} and \eqref{reprule} we find the following
explicit result for the off-shell free energy of the RB theory:
\begin{align}\label{Feff}
F_B(c_B, V) &= \frac{N_B}{6\pi} {\Bigg[}  -
\frac{ \left(\lambda_B-2{\rm sgn}(\lambda_B) \right) }{\lambda_B}\hat{c}_B^3 - \frac{8\Lambda}{|\lambda_B|^3}(2\pi {\hat V}^2)^3\nonumber\\
&\qquad -\frac{3}{|\lambda_B|}  (\hat{c}_B^2 - {\hat m}_B^2)2\pi {\hat V}^2  +  \frac{6 {\hat b}_4}{\lambda_B}(2\pi{\hat V}^2)^2 + \frac{ ( 3x^B_6 + 3)}{|\lambda_B|}(2\pi {\hat V}^2)^3\nonumber\\
&\qquad+3 \int_{-\pi}^{\pi} \rho_B(\alpha) d\alpha\int_{\hat{c}_B}^{\infty}dy y\left(\ln\left(1-e^{-y-i\alpha}\right)+\ln\left(1-e^{-y+i\alpha}\right)  \right)
{\Bigg]}\ .
\end{align}
We compare the result \eqref{Feff} with the prediction
\eqref{ferdualmintro} of duality for the Higgsed phase free energy
given by
\begin{align}\label{Feff1}
  F_B(c_B, V)
& = \frac{N_B }{6\pi} \Bigg[- \frac{\lambda_B - 2\sgn(\lambda_B)}{\lambda_B} {\hat c}_B^3 \nonumber\\
&\qquad - \frac{3 }{|\lambda_B|}  (\hat{c}_B^2 - \hat{m}_B^2) 2\pi \hat{V}^2   +\frac{6  \hat b_4}{\lambda_B}(2\pi \hat{V}^2)^2 + \frac{(3  x^B_6 + 4)}{|\lambda_B|} (2 \pi \hat{V}^2)^3 \nonumber \\  
&\qquad + 3 \int_{-\pi}^{\pi}d\alpha \rho_{B}(\alpha)\int_{{\hat c}_B}^{\infty} dy ~y~\left( \ln\left(1-e^{-y-i\alpha}\right)  + \ln\left(1-e^{-y+i\alpha}\right)\right)\Bigg]\ ,
\end{align}
where, to get the above expression, we have used \eqref{unitary} and
\eqref{mdefnintro} to write the field $\sigma_B$ in
\eqref{ferdualmintro} as
\begin{equation}\label{mdef}
 \sigma_B = \frac{2\pi V^2}{|\lambda_B|}\ .
\end{equation}

We see that \eqref{Feff} agrees precisely with \eqref{Feff1}
provided we choose the as yet undetermined parameter $\Lambda$ as  
\begin{equation}\label{alphapred}
\Lambda =  -\frac{1}{8}\lambda_B^2\ .
\end{equation}

In the next subsection we will verify that the result
\eqref{alphapred} - which is so far just a prediction of duality - can
also be obtained by direct computation within the Higgsed CB
theory. The strategy we employ is the following.  We first note that
the gap equation corresponding to stationarity of \eqref{Feff} with
respect to ${\hat V}^2$ is
\begin{equation}\label{gappred}
{\hat c}_B^2 - {\hat m}_B^2 + 8\Lambda {\hat \sigma}_B^2 - 4\hat b_4 \lambda_B {\hat \sigma}_B - (3 x_6^B + 3)\lambda_B^2 {\hat \sigma}_B^2 = 0\ .
\end{equation}
where $\sigma_B$ is given in terms of $V^2$ by \eqref{mdef}.  The
equation \eqref{gappred} merely simply expresses the condition that
the tadpole of the fluctuation of the scalar field $V$ vanishes when
the field $V$ is expanded around its true solution $v$. In the next
subsection we directly evaluate this `tadpole vanishing condition' in
the RB theory in the Higgsed phase and thereby determine $\Lambda$ by
comparison with \eqref{gappred}.

\subsection{Tadpole cancellation for $V$}
As we have explained above, in the Higgsed phase our scalar field $V$
gets the expectation value $v$. It is useful to define
\begin{equation} \label{deffh}
V(x)= v + H(x)\ .
\end{equation}
The condition that $v$ is the correct vacuum expectation value of
$V(x)$ is equivalent to the condition that the expectation value
(i.e. one point function, i.e. tadpole) of the fluctuation $H(x)$
vanishes. In other words we require that
\begin{equation}\label{nseffH}
  \int_{\mbb{R}^2 \times S^1} 
  [dV dW dZ dA]\  H(x)\ e^{-S_{\text{E}}[A,W,Z,V]} =0\ .
\end{equation}
Using the explicit form of $S_{\text{E}}[A,W,Z,V]$ in \eqref{asb},
equation \eqref{nseffH} can be rewritten as
\begin{equation} \label{exptc} {\rm sgn}(\kappa_B) \langle
  \bar{W}_{a\mu}(x) W^{a\mu}(x) \rangle + |\kappa_B| \langle Z_\mu(x)
  Z^\mu(x) \rangle \ + \frac{\partial } {\partial (v^2)}U_{\rm cl}(v^2) =0\ .
\end{equation} 
where $U_{\rm cl}(V)$ is the potential for the Higgs field $V$ given in
\eqref{vpot} and all expectation values are evaluated about the
`vacuum' where $V(x)=v$. While the first and third terms in
\eqref{exptc} above are both of order $N_B$, it is easily verified
that the second term in this equation - the term proportional to
$\langle Z_\mu (x)^2 \rangle $ is of order unity\footnote{This follows
  from the observation that the $Z$ propagator scales like $1/N_B$.}
and so can be dropped in the large $N_B$ limit. At leading order in
the large $N_B$ limit, it follows that the tadpole cancellation
condition \eqref{nseffH} can be rewritten as
\begin{equation} \label{exptcn} 
\frac{\lambda_B}{2 \pi \cV_3} \int d^3 x \langle \bar{W}_{a\mu}(x)  W^{a\mu}(x) \rangle  + \frac{\partial U_{\rm cl}(\sigma_B)} {\partial \sigma_B} =0\ ,
\end{equation} 
where we have integrated the equation \eqref{exptc} over spacetime and
have divided the resulting expression by the volume of spacetime
$\cV_3$. We have also changed variables to
$\sigma_B = 2\pi v^2 / |\lambda_B|$ defined in \eqref{mdef}.  Moving to
momentum space, \eqref{exptcn} turns into
\begin{equation} \label{vgap} 
\frac{\lambda_B}{2 \pi} \int \frac{\td^3 p}{(2\pi)^3} \eta^{\mu\nu}\,G^a_{a\mu\nu}(p)  + \frac{\partial U_{\rm cl}(\sigma_B)} {\partial \sigma_B} =0\ ,
\end{equation}
where 
\begin{equation} \label{propmomn}
\langle \bar{W}_{a\mu}(-p) W^b_\nu (p')\rangle= 
G^b_{a\mu\nu}(p)\,  (2 \pi)^{3} \delta^{(3)}(p-p') = \delta_a{}^b\ G_{\mu\nu}(p)\,  (2 \pi)^{3} \delta^{(3)}(p-p')\ ,
\end{equation}
and the measure $\td$ is the natural measure in momentum space at
finite temperature.\footnote{Explicitly, the notation $\td^3 p$
  signifies that we work at finite temperature i.e. on the spacetime
  $\mbb{R}^2 \times S^1_\beta$ where the third direction $x^3$ is a
  circle of circumference $\beta$. The measure $\td^3 p$ is then given
  by
  \begin{equation}
    \int \frac{\td^3 p}{(2\pi)^3} f(p) = \int \frac{dp_1 dp_2}{(2\pi)^2} \int_{-\pi}^\pi d\alpha\, \rho_B(\alpha) \frac{1}{\beta} \sum_{n=-\infty}^\infty f\left(\frac{2\pi n + \alpha}{\beta}\right)\ ,
  \end{equation}
  where $\rho_B(\alpha)$ is the distribution of the eigenvalues
  $\alpha$ of the gauge field holonomy around $S^1_\beta$.}  Happily,
the exact all-orders formula for the propagator $G_{\mu\nu}$ was
computed in \cite{Choudhury:2018iwf}. In Appendix \ref{app} we proceed
to plug the explicit expression for $G_{\mu\nu}$ and evaluate the
first term in \eqref{vgap}. We are able to evaluate all the relevant
summations and integrals analytically, and demonstrate that in our
choice of regularisation scheme \eqref{exptcn} takes the explicit form
\begin{equation}\label{vgapsim}
-\frac{N_B}{2\pi} \left({ c}_B^2-\lambda_B^2{ \sigma}_B^2 \right) +  \frac{\partial U_{\rm cl}(\sigma_B)}{\partial { \sigma}_B} = 0\ .
\end{equation}
Recall the expression for $U(\sigma_B)$ from \eqref{vpot}:
\begin{align}
  U_{\rm cl}(\sigma_B) &= \frac{N_B}{2\pi} \left( m_B^2 \sigma_B   + 2 b_4 \lambda_B \sigma_B^2 +  ( x^B_6+1) \lambda_B^2 \sigma_B^3\right)\ .
\end{align}
Plugging this back into \eqref{vgapsim} we find 
\begin{equation}\label{vgapfinal}
	{ c}_B^2 - { m}_B^2 - 4 {b}_4 \lambda_B {\sigma}_B - (3x_6^B + 4) \lambda_B^2 {\sigma}_B^2= 0\ ,
\end{equation}
Comparing this with the gap equation obtained earlier in
\eqref{gappred}, we see that \eqref{gappred} matches \eqref{vgapfinal}
for the predicted value of $\Lambda = -\lambda_B^2 / 8$ in
\eqref{alphapred} as expected.

\section{A three variable off-shell free energy} \label{ose} 

The finite temperature unHiggsed phase is governed by the two-variable
off-shell free energy (equation \eqref{bosoff} in Appendix
\ref{review})
\begin{align}\label{bosoffmain}
F_B(c_B, \tl\cS) &= \frac{N_B}{6\pi} \Bigg[ -{\hat c}_B^3+ 3 {\tilde \cS} \left({\hat c}_B^2-{\hat m}_B^2 \right) + 6 {\hat b_4} \lambda_B  {\tilde \cS}^2 - (4+3 x_6^B)\lambda_B^2 {\tilde \cS}^3\nonumber \\
&\qquad+ 3 \int_{-\pi}^{\pi} d\alpha \rho_{B}(\alpha)\int_{{\hat c}_B}^{\infty} dy ~y~\left( \ln\left(1-e^{-y-i\alpha}\right)  + \ln\left(1-e^{-y+i\alpha}\right)\right)\Bigg]\ ,
\end{align}
On the other hand, we have demonstrated in this paper that the finite
temperature Higgsed phase is governed by the two-variable off-shell
free energy \eqref{ferdualmintro}. As these are two separate `phases'
of the same theory it is somewhat unsatisfying that the off-shell
`Landau-Ginzburg' free energies used to describe them are
different. The reader may wonder whether there exists a single master
off-shell free energy functional - analytic in all `fields' - which
encompasses the physics of both \eqref{bosoffmain} and
\eqref{ferdualmintro}. At least at the algebraic level there is a simple
affirmative answer to this question as we now describe.

Consider the off-shell free energy
\begin{align}\label{Feffofsh}
F(c_B, \sigma_B, \tl\cS) & = \frac{N_B }{6\pi} \Bigg[- {\hat c}_B^3- 4  {\tilde\cS}^3 \lambda_B^2-3 {\hat c}_B^2  {\hat \sigma}_B-12{\tilde\cS}^2 \lambda_B^2 {\hat \sigma}_B -12{\tilde\cS} \lambda_B^2 {\hat \sigma}_B^2  \nonumber\\
&\qquad\qquad + 6 {\hat c}_B |\lambda_B| ({\tilde\cS} + {\hat \sigma}_B)^2 + 3  \left({\hat m}_B^2 {\hat \sigma}_B + 2\lambda_B   {\hat b}_4 {\hat \sigma}_B^2 + \lambda_B^2 x_6^B {\hat \sigma}_B^3\right) \nonumber\\
 &\qquad\qquad + 3 \int_{-\pi}^{\pi}d\alpha \rho_{B}(\alpha)\int_{{\hat c}_B}^{\infty} dy ~y~\left( \ln\left(1-e^{-y-i\alpha}\right)  + \ln\left(1-e^{-y+i\alpha}\right)\right)\Bigg].
\end{align}
Note that \eqref{Feffofsh} is a function of three `field' variables,
namely $c_B$, ${\tilde\cS}$ and $\sigma_B$. Extremizing
\eqref{Feffofsh} w.r.t. ${\tilde\cS}$, $c_B$ and $\sigma_B$
respectively yields the equations
\begin{align} \label{foll}
&({\tilde\cS}+{\hat \sigma}_B) (-{\hat c}_B + |\lambda_B| ({\tilde\cS}+{\hat \sigma}_B))=0\ ,\nonumber\\
&{\hat c}_B (\cS(c_B)+{\hat \sigma}_B)-|\lambda_B| ({\tilde\cS}+{\hat \sigma}_B)^2=0\ ,\nonumber\\
&{\hat c}_B^2 - {\hat m}_B^2-4 {\hat c}_B |\lambda_B|  ({\tilde\cS}+{\hat \sigma}_B) +\lambda_B \left(4 {\tilde\cS}^2 \lambda_B- 4 {\hat b}_4 {\hat \sigma}_B+ 8 \lambda_B {\hat \sigma}_B {\tilde\cS}-3 \lambda_B {\hat \sigma}_B^2 x_6^B \right)=0\ .
\end{align}
The quantity $\cS(c_B)$ that appears in the second of \eqref{foll} is
defined in \eqref{nss}. Off-shell, the objects $\cS(c_B)$ and
${\tilde \cS}$ are completely distinct.  $\cS(c_B)$ is a
function of $c_B$ while ${\tilde \cS}$ is an independent
variable. However it is easy to see (by subtracting the first two
equations in \eqref{foll}) that these two quantities are, in fact,
equal on-shell.

Note in particular that the first of \eqref{foll} - the equation that
follows upon extremizing \eqref{Feffofsh} w.r.t. ${\tilde\cS}$ - is
the product of two factors. This equation is satisfied either if
\begin{equation}
\label{foee}
{\tilde\cS}+{\hat \sigma}_B=0\ ,
\end{equation}
or if 
\begin{equation}
\label{soee}
 -{\hat c}_B + |\lambda_B| ({\tilde\cS}+{\hat \sigma}_B)=0\ .
\end{equation}
(clearly \eqref{foee} and \eqref{soee} cannot simultaneously be obeyed
unless $c_B=0$).  Let us first suppose that \eqref{foee} is
obeyed. Using \eqref{foee} to eliminate ${ \sigma}_B$ from
\eqref{Feffofsh} yields an off-shell free energy that now depends only
on ${\tilde\cS}$ and $c_B$. It is easily verified that the resultant
free energy agrees exactly with the two-variable free energy
\eqref{bosoffmain} in the unHiggsed phase. It follows that solutions
of \eqref{foee} parametrize - and govern the physics of - the
unHiggsed phase of the RB theory.

In a similar manner let us now suppose that \eqref{soee} is obeyed in
which case we use it to eliminate ${\tilde\cS}$. It is easily verified
that the resultant two-variable free energy - which depends on $c_B$
and ${ \sigma}_B$ - agrees exactly with \eqref{Feff1} with the
identification \eqref{mdef}:
\begin{equation}\label{frd}
\sigma_B = \frac{2\pi v^2}{|\lambda_B|}\ .
\end{equation}
It follows that solutions of \eqref{soee} parametrize - and govern the
physics of - the Higgsed phase of the RB theory.

The identification \eqref{frd} has a simple explanation. Recall that
the bare mass $m_B^2$ appeared in the action \eqref{rst} as the
coefficient of ${\bar \phi} \phi$. It follows that the Legendre
transform of the free energy of our theory w.r.t.~$m_B^2$
yields the exact quantum corrected effective potential of our theory
as a function of the composite field
$( {\bar \phi} \phi )_{\rm cl}$.  This Legendre transform may
be computed by adding the term
$$-m_B^2 ( {\bar \phi} \phi )_{\rm cl}$$ to \eqref{Feffofsh} and
then treating $m_B^2$ as a new dynamical field w.r.t.~which
\eqref{Feffofsh} has to be extremized (of course we also continue to
extremize \eqref{Feffofsh} w.r.t. $c_B$, $\sigma_B$ and ${\tilde \cS}$
as before). Note that the dependence of \eqref{Feffofsh} on $m_B^2$ is
extremely simple; it occurs entirely through the term
$\tfrac{N_B}{2 \pi} \sigma_B m_B^2$. As a consequence, extremizing
w.r.t. $m_B^2$ sets
\begin{equation} \label{frdnn} 
{ \sigma}_B= \frac{2 \pi ( {\bar \phi} \phi )_{\rm cl}  }{N_B}\ .
\end{equation}
In the Higgsed phase it follows from
$\phi^i = \delta^{i N_B} \sqrt{|\kappa_B|}\, v$ (equation
\eqref{unitary}) that
\begin{equation}
\label{neh} 
( {\bar \phi} \phi )_{\rm cl}= |\kappa_B| v^2\ .
\end{equation}
(in obtaining \eqref{neh} we use the fact that the Higgs field $V$ is
effectively classical in the large $N_B$ limit).  Inserting
\eqref{neh} into \eqref{frdnn} yields \eqref{frd}.  We note, however,
that \eqref{frdnn} is more general than \eqref{frd} because it applies
even in the unHiggsed phase.  We will make use of this fact in the
next section.

We have thus found a simple single off-shell free energy - namely
\eqref{Feffofsh} - that captures the physics of both the Higgsed and
the unHiggsed phases. We have also explained that one of the three
variables that appears in this free energy - namely $\sigma_B $ - has
a simple direct physical interpretation given by \eqref{frdnn}.  It
follows, in particular, that if we integrate $c_B$ and ${\tilde \cS}$
out from \eqref{Feffofsh}, the resultant free energy (which is a
function of $\sigma_B$) can be reinterpreted as the quantum effective
potential of the theory as a function of $ ({\bar \phi} \phi )_{\rm cl}$.
In the next section we will explicitly undertake this exercise in the
zero temperature limit.

It is easily verified that the duality map \eqref{dualitymapmain}
between parameters together with the field redefinitions
\begin{equation} \begin{split}
\label{dmfr}
&\lambda_B {\tilde \cS} = \lambda_F {\tilde \cC} - \frac{\sgn(\lambda_F)}{2}{\hat c}_F\ ,\quad \lambda_B{ \sigma}_B= -\frac{2 \pi \zeta_F}{\kappa_F }\ ,\quad c_B= c_F\ .
\end{split}
\end{equation}
turns the bosonic off-shell free energy \eqref{Feffofsh} into the
fermionic off-shell free energy \eqref{ofcrf12}. This match captures
the Bose Fermi duality between RB and CF theories at the level of the
complete thermal off-shell free energies of the two theories; note
that each of these off-shell free energies is analytic in all
`fields'.  In Appendix \ref{cbl} we investigate the behaviour of our
three-variable off-shell free energy in the so called critical boson
scaling limit of the RB theory.

\section{The exact Landau-Ginzburg effective potential} \label{ost} In
this section we integrate out the variables ${\tilde \cS}$ and $c_B$
out from the effective action \eqref{bosoffmain} and obtain an
off-shell free energy for the field ${\sigma}_B$. We work at zero
temperature throughout this section. In this simple - and
physically especially important - limit we obtain a simple analytic
expression for the resultant free energy as a function of
$\sigma_B$. As we have explained in the previous section, this free
energy is simply related to the quantum effective potential of the RB
theory as a function of the field $({\bar \phi}\phi)_{\rm cl}$.

After having obtained this exact Landau-Ginzburg potential we study
and use it in various ways. First we note that this effective
potential has extrema of two sorts - local maxima and local
minima. Local maxima represent unstable saddle point solutions of the
theory. In the case of the unHiggsed branch (see below) we present an
interpretation of the resultant instability in terms of the tachyonic
bound states of the system.  We also use the exact Landau-Ginzburg
effective action that we obtain to understand the zero temperature
phase diagram of the RB theory (as a function of its microscopic
parameters) in a simple and intuitive way. Finally we also make a
prediction for the range of the parameter $x_6$ over which the RB
theory is stable, i.e. has a stable vacuum.

\subsection{An effective potential for ${ \sigma}_B$}
In the zero temperature limit the three-variable off-shell free energy
\eqref{Feffofsh} simplifies to
\begin{align}
F(c_B, \sigma_B, \tl\cS) & = \frac{N_B }{6\pi} \Big[- {\hat c}_B^3- 4  {\tilde\cS}^3 \lambda_B^2-3 {\hat c}_B^2 {\hat \sigma}_B-12{\tilde\cS}^2 \lambda_B^2 {\hat \sigma}_B -12{\tilde\cS} \lambda_B^2 {\hat \sigma}_B^2   \nonumber\\
&\qquad\qquad +6 {\hat c}_B |\lambda_B| ({\tilde\cS}+{\hat \sigma}_B)^2 + 3 \left({\hat m}_B^2 {\hat \sigma}_B  +2\lambda_B {\hat b}_4 {\hat \sigma}_B^2 + x_6^B \lambda_B^2 {\hat \sigma}_B^3\right) \Big]\ .
\end{align}
Varying this free energy w.r.t. ${\tilde \cS}$ produces the first of
the gap equations in \eqref{foll} which we repeat here for convenience
\begin{align}
({\tilde\cS}+{\hat \sigma}_B) (-{\hat c}_B + |\lambda_B| ({\tilde\cS}+{\hat \sigma}_B))=0\ .
\end{align}
As we have discussed above, this equation has two solutions
corresponding to the unHiggsed and Higgsed branches:
\begin{align}
  \text{unHiggsed}:&\quad \tl\cS = -{\hat \sigma}_B\ ,\qquad  \text{Higgsed}:\quad \tl\cS = -{\hat \sigma}_B + \frac{{\hat c}_B}{|\lambda_B|}\ .
\end{align}
Plugging these solutions back into the expression for the free energy,
we have, in the unHiggsed phase,
\begin{align}\label{Fuhig}
F^{({\rm uH})}(c_B, \sigma_B) & = \frac{N_B }{6\pi T^3} \left(-{ c}_B^3 + 4 \lambda_B^2{ \sigma}_B^3 - 3({ c}_B^2 - { m}_B^2) { \sigma}_B   +  6 { b}_4 \lambda_B { \sigma}_B^2   + 3 x_6^B   \lambda_B^2{ \sigma}_B^3\right)\ ,
\end{align}
and in the Higgsed phase,
\begin{align}\label{Fhig}
F^{({\rm H})}(c_B, \sigma_B) & = \frac{N_B }{6\pi T^3} \left(\tfrac{2 - |\lambda_B|}{|\lambda_B|} { c}_B^3 +  4\lambda_B^2{ \sigma}_B^3    - 3({ c}_B^2 -{ m}_B^2) { \sigma}_B   + 6 { b}_4 \lambda_B { \sigma}_B^2   + 3   x_6^B \lambda_B^2{ \sigma}_B^3 \right)\ .
\end{align}
We then extremize the above free energies with respect to $c_B$ to get
\begin{align} \label{solssig}
 \text{unHiggsed}:&\quad { c}_B =
  -2{ \sigma}_B\ ,\qquad \text{Higgsed}:\quad { c}_B =
  \frac{|\lambda_B|}{2 - |\lambda_B|} 2{ \sigma}_B\ .
\end{align}
Recall that $c_B$ is positive by definition. It follows that the
solutions \eqref{solssig} exist only when ${\sigma}_B$ is positive
(negative) in the Higgsed (unHiggsed) phase respectively. Plugging
back the above expressions into the free energies in \eqref{Fuhig} and
\eqref{Fhig}, we get
\begin{align}\label{Fuhigfin}
F^{({\rm uH})}(\sigma_B) & = \frac{N_B }{2\pi T^3} \left[\left((1 + x_6^B) -  \frac{4-\lambda_B^2}{3\lambda_B^2}  \right) \lambda_B^2 { \sigma}_B^3 +  2 { b}_4 \lambda_B  { \sigma}_B^2 + { m}_B^2 { \sigma}_B \right]\ ,
\end{align}
and in the Higgsed phase,
\begin{align}\label{Fhigfin}
F^{({\rm H})}(\sigma_B) & = \frac{N_B }{2\pi T^3} \left[\left((1 +  x_6^B) - \frac{|\lambda_B|(4 - |\lambda_B|)}{3(2-|\lambda_B|)^2}\right)\lambda_B^2 { \sigma}_B^3 +  2 { b}_4 \lambda_B { \sigma}_B^2 + { m}_B^2 { \sigma}_B \right]\ .
\end{align}
The quantum effective potential for the field
$(\bar\phi \phi)_{\rm cl}$ is related to the above free energies as
\begin{equation}
  U_{\rm eff}((\bar\phi\phi)_{\rm cl}) = T^3 F(\sigma_B)\quad\text{with the replacement}\quad \sigma_B \to \frac{2\pi (\bar\phi\phi)_{\rm cl}}{N_B}\ .
\end{equation}
We continue to use the variable $\sigma_B$ as the argument of the
effective potential $U_{\rm eff}$ to avoid clutter, with the
understanding that all instances of $\sigma_B$ in $U_{\rm eff}$ are to
be replaced with $2\pi (\bar\phi \phi)_{\rm cl} / N_B$. Explicitly, we
have
\begin{empheq}[box=\fbox]{align}\label{LGpot}
U_{\rm eff}(\sigma_B) &= \left\{\arraycolsep=1.4pt\def\arraystretch{2.2}\begin{array}{cl} \frac{N_B}{2 \pi} 
  \left[ (x_6 - \phi_2) \lambda_B^2 { \sigma}_B^3 + 2 \lambda_B b_4 { \sigma}_B^2  +  m_B^2 { \sigma}_B \right] & \quad\text{for }\sigma_B < 0\ , \\ \frac{N_B}{2\pi}\left[( x_6 - \phi_1)\lambda_B^2 { \sigma}_B^3 +  2 \lambda_B{ b}_4  { \sigma}_B^2 + { m}_B^2 { \sigma}_B\right]  & \quad\text{for }\sigma_B > 0\ , \end{array}\right.\nonumber\\
&\qquad\ \text{with the replacement}\quad \sigma_B \to \frac{2\pi (\bar\phi \phi)_{\rm cl}}{N_B}\ .
\end{empheq}
The constants $\phi_1$ and $\phi_2$ are given by
\begin{equation}\label{phi12def}
\phi_1 = \frac{4}{3}\left(\frac{1}{(2-|\lambda_B|)^2} - 1\right)\ ,\quad \phi_2 = \frac{4}{3}\left(\frac{1}{\lambda_B^2} - 1\right)\ .
\end{equation}
Observe that the effective potential \eqref{LGpot} is bounded from
below for positive values of $\sigma_B$ if the coefficient of
$\sigma_B^3$ is positive in the second of \eqref{LGpot}, i.e. when
\begin{equation}\label{phi1def}
x_6 > \phi_1\ .
\end{equation}
Similarly, the effective potential is bounded from below for negative
values of $\sigma_B$ if the coefficient of the $\sigma_B^3$ term is
negative in the first of \eqref{LGpot}, i.e. when
\begin{equation}\label{phi2def}
x_6 < \phi_2\ .
\end{equation}
Note that $\phi_1 < \phi_2$. 

Note that the terms proportional to $\sigma_B^2$ and $\sigma_B$ are
identical for the two ranges of $\sigma_B$ but the coefficients of the
$\sigma_B^3$ terms are different: this non-analyticity in the cubic
term is what gives a sharp distinction between the Higgsed and
unHiggsed branches of the effective potential at zero temperature.
When we turn on temperature we expect this non-analyticity to be
smoothed out. \footnote{This should be easy to verify - and seems to
  follow from the fact that the finite temperature free energy is an
  analytic function of its variables - but we have not verified it in
  detail.}

We also give a slightly different expression for the Landau-Ginzburg
potential in terms of the variable $c_B$ which is useful for the
analysis of the gap equations as performed in Section 4 of
\cite{abcd}. For this purpose, we substitute back the expressions for
$\sigma_B$ in terms of $c_B$ from \eqref{solssig}:
\begin{align}\label{gappot}
&U_{\rm eff}(c_B) = \left\{\arraycolsep=1.4pt\def\arraystretch{2.2}\begin{array}{cl} \frac{N_B}{2 \pi} 
  \left[ A_u \frac{ { c}_B^3}{6} +  B_{4, u} \frac{{ c}_B^2}{2}  -  m_B^2 \frac{{ c}_B}{2} \right] & \quad\text{unHiggsed}\ , \\ \frac{N_B(2-|\lambda_B|)}{2\pi|\lambda_B|}\left[-A_h  \frac{ c_B^3}{6} -  B_{4,h} \frac{c_B^2}{2} + m_B^2 \frac{c_B}{2}\right]  & \quad\text{Higgsed }\ . \end{array}\right.
\end{align}
 Here, $A_u$, $B_{4,u}$ and
$A_h$, $B_{4,h}$ are constants defined by
\begin{alignat}{2}\label{abdef}
&A_u= 1-\left( 1+ \frac{3 x_6}{4} \right) \lambda_B^2\ ,\quad
&& B_{4,u}= \lambda_B b_4 \ ,\nonumber\\
&A_h=1-  \left( 1+\frac{3x_6}{4}\right) (2-|\lambda_B|)^2\ ,\quad  &&{B}_{4,h} =  - \sgn(\lambda_B)(2-|\lambda_B|) b_4\ .
\end{alignat}
The off-shell variable $c_B$ has the following advantage; its on-shell
value coincides with the pole mass (the \emph{gap}) of the fundamental
excitation in the corresponding phase. We record the gap equations
that follow from extremizing \eqref{gappot} w.r.t~the variable $c_B$:
\begin{align}\label{exactgap}
\text{unHiggsed}:&\quad A_u c_B^2 + 2 B_{4,u} c_B - m_B^2 = 0\ ,\nonumber\\
  \text{Higgsed}:&\quad A_h c_B^2 + 2 B_{4,h} c_B - m_B^2 = 0\ .
\end{align}
Solutions to the quadratic equations in \eqref{exactgap} above
correspond to candidates for the Higgsed or unHiggsed phases of the
theory. The very recent paper \cite{abcd} analysed the solutions of
\eqref{exactgap} in detail and used information about the free energy
at these solutions to obtain the phase structure as a function of the
parameters $x_6$, $\lambda_B b_4$ and $m_B^2$. In the next subsection
we will extract the same physical information from the exact
Landau-Ginzburg effective potential \eqref{LGpot}.

\subsection{Higgsed branch} \label{hp}

In this subsection we study the effective potential \eqref{LGpot}
in more detail on the Higgsed branch. 

\subsubsection{Potential for the Higgs vev}

We have already noted around \eqref{frd} that the variable $\sigma_B$
has a simple interpretation in terms of the Higgs vev on the Higgsed
branch. Making the replacement \eqref{frd}, i.e.
$\sigma_B = 2\pi v^2 / |\lambda_B|$ in the second line of
\eqref{LGpot} we find
\begin{align}\label{Feff1mm}
U_{\rm eff}(v)
& = \frac{N_B}{|\lambda_B|}\left( m_B^2 v^2 + 4\pi \sgn(\lambda_B) b_4 v^4 + \left( (1 + x_6^B) - \tfrac{|\lambda_B|(4-|\lambda_B|)}{3(2 - |\lambda_B|)^2}\right) 4 \pi^2   v^6\right)\ ,
\end{align}
It is easily verified that \eqref{Feff1mm} may also be obtained by
taking the zero temperature limit of \eqref{Feff1} (i.e. dropping the
last line in that formula) and integrating $c_B$ out of that
equation. It follows that the true value of the Higgs vev in the
vacuum is obtained by extremizing \eqref{Feff1mm}.

Note that \eqref{Feff1mm} manifestly reduces to the classical potential
\begin{equation}\label{classpot}
U_{\rm cl}(v) = \frac{N_B}{|\lambda_B|}
\left( m_B^2 v^2 + 4 \pi\sgn(\lambda_B) b_4 v^4 + (2 \pi)^2 (x_6+1) v^6 \right)\ ,
\end{equation}
in the classical limit \eqref{class} (i.e.~$\lambda_B \to 0$ with
$m_B^2$, $b_4$ and $x_6$ fixed) as expected on general
grounds. 

\subsubsection{Graphs of $U_{\rm eff}(\sigma_B)$ in various cases}

In this subsubsection we will study the graphs of
$U_{\rm eff}(\sigma_B)$ on the Higgsed branch for various ranges of
values of microscopic parameters. The results of this subsubsection
will prove useful in sketching the phase diagram of the RB theory in
later subsections.

Recall that, on the Higgsed branch, $U_{\rm eff}(\sigma_B)$ is given
by the expression
\begin{equation}\label{ueffsb}
U_{\rm eff}(\sigma_B) = \frac{ N_B }{2\pi}\left(-  \tfrac{|\lambda_B|^2}{(2-|\lambda_B|)^2}  {A_h}\frac{4 \sigma_B^3}{3} - \tfrac{|\lambda_B|}{(2-|\lambda_B|)} 2{B}_{4, h}  \sigma_B^2 + m_B^2 \sigma_B  \right)\ ,\quad \sigma_B > 0\ ,
\end{equation}
($A_h$, $B_{4,h}$ were defined in the second line of \eqref{abdef}).
The extremization of $U_{\rm eff}(\sigma_B)$ produces the gap equation
in the second of \eqref{exactgap} which we reproduce here:
\begin{equation}\label{hatc}
 {A_h} c_B^2 + 2 { B_{4,h}} c_B -m_B^2=0\ .
\end{equation}

The structure of the curves for $U_{\rm eff}$ above turn out to depend
sensitively on discriminant $D_h$ of this gap equation \eqref{hatc}:
\begin{equation}\label{disch}
D_h = 4( B_{4,h}^2 + m_B^2 A_h)\ .
\end{equation}

As discussed below \eqref{LGpot}, the above effective
potential is bounded below when $x_6> \phi_1$ where $\phi_1$ was
defined in \eqref{phi12def}. In other words  $A_h<0$ when $x_6 > \phi_1$ and $A_h>0 $ when $x_6< \phi_1$. 
\footnote{Note that $\phi_1$ is an increasing function of
$\lambda_B$. In particular $\phi_1= -1$ for the free theory
($\lambda_B=0$), whereas for the strongly coupled case
($|\lambda_B|=1$) we have $\phi_1=0$. The fact that $\phi_1$ increases
as we increase $|\lambda_B|$ indicates that coupling effects increase
the propensity of our theory to develop a runaway instability along the
$v$ direction.}

We have the following cases as depicted in Figures \ref{higneg} and \ref{higpos}:
\begin{enumerate}
\item \emph{$A_{h}$ is negative}: the potential $U_{\rm eff}$
  increases at large $\sigma_B$.
      \begin{figure}[htbp]
        \begin{center}
          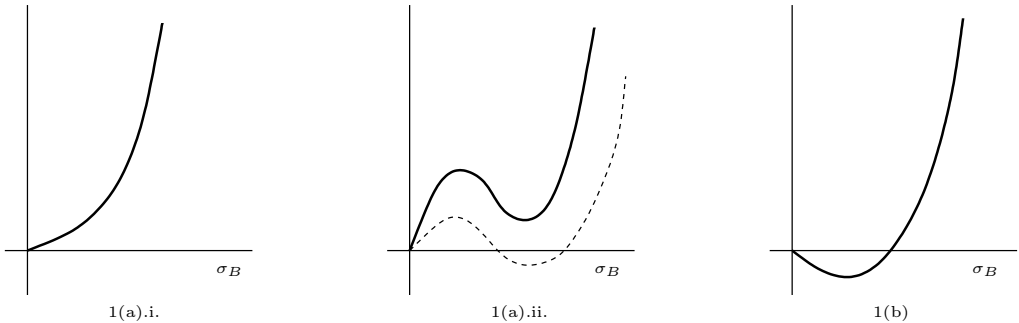
          \caption{Effective potential in the Higgsed phase for $A_h$ negative.}
          \label{higneg}
        \end{center}
      \end{figure}
  \begin{enumerate}
  \item \emph{$m_B^2$ is positive}:
    \begin{enumerate}
    \item \emph{$B_{4, h}$ is negative, or $B_{4,h}$ is positive
        such $D_h$ is negative }: the potential rises monotonically as
      $\sigma_B$ increases from zero to infinity, and there are no nontrivial
      positive solutions of the gap equation \eqref{hatc}.

    \item \emph{$B_{4, h}$ is positive such that $D_h$ is
        positive}: As $\sigma_B$ is increased from zero, $U_{\rm eff}$
      initially increases, reaches a local maximum and then decreases,
      reaches a minimum and then increases without bound. In this case
      the gap equation has two solutions; the larger of which is the
      candidate for a stable phase (the smaller solution presumably
      describes unstable dynamics since it occurs at a local maximum
      of the effective potential).
    \end{enumerate}
  \item \emph{$m_B^2$ is negative}: For either sign of $B_{4, h}$,
    $U_{\rm eff}$ initially decreases, reaches a minimum and then
    turns and increases indefinitely. The gap equation has exactly one
    legal solution (i.e.~a solution for $c_B$ which is positive) which
    is the candidate for a stable phase.
  \end{enumerate}
\item \emph{$A_{h}$ is positive}: the potential $U_{\rm eff}$
  decreases at large $\sigma_B$.
      \begin{figure}[htbp]
        \begin{center}
          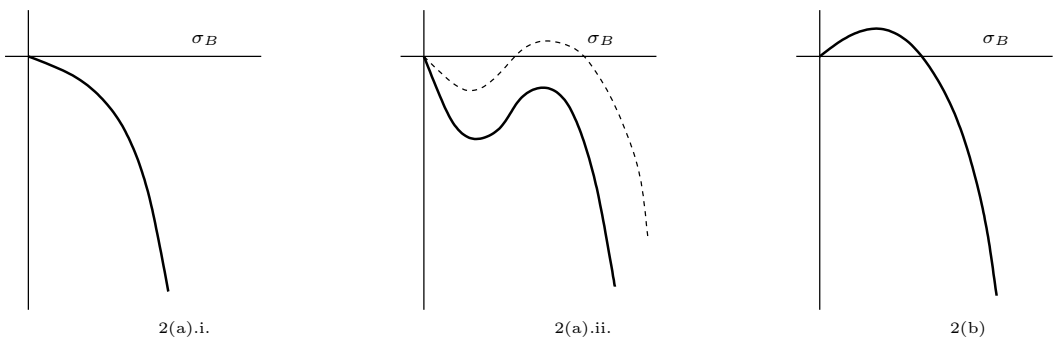
          \caption{Effective potential in the Higgsed phase for $A_h$ negative.}
          \label{higpos}
        \end{center}
      \end{figure}
  \begin{enumerate}
  \item \emph{$m_B^2$ is negative}
    \begin{enumerate}
    \item \emph{$B_{4, h}$ is positive, or $B_{4, h}$ is negative
        such that $D_h$ is negative}: The potential decreases
      monotonically as $\sigma_B$ increases from zero to infinity, and
      there are no nontrivial positive solutions of the gap equation
      \eqref{hatc}.
    \item \emph{$B_{4, h}$ is negative such that $D_h$ is
        positive}: As $\sigma_B$ is increased from zero $U_{\rm eff}$
      initially decreases, reaches a local minimum and then increases
      till it reaches a local maximum after which it decreases without
      bound. In this case the gap equation has two solutions; the
      smaller of which is the candidate metastable phase (the larger
      solution presumably describes unstable dynamics again since it
      occurs at a local maximum of the potential).
    \end{enumerate}
  \item \emph{$m_B^2$ is positive}: For either sign of $B_{4, h}$,
    $U_{\rm eff}$ initially increases, reaches a maximum and then
    turns and decreases indefinitely. The gap equation has exactly one
    legal solution; this is a local maximum and so presumably
    describes an unstable `phase'.
  \end{enumerate}
\end{enumerate}
In the last two paragraphs above we have encountered three examples of
solutions (1.a.ii, 2.a.ii, 2.b) to the gap equations that describe
unstable `phases'. For future use we note that these three unstable
solutions are all given by the following root of \eqref{hatc}:
\begin{equation}\label{roeeuns}
c_B= \frac{-B_{4, h}+\sqrt{B_{4,h}^2+A_h m_B^2}}{A_{h}}\ .
\end{equation}
It is also easy to check that the three cases described above are the
only three legal roots of the form \eqref{roeeuns}.  In other words
every local maximum of the potential \eqref{Feff1mm} is a root of the
form \eqref{roeeuns}, and a legal root (i.e.~a root for which the RHS
is positive) of the form \eqref{roeeuns} is one of the three `local
maxima' situations described above \footnote{To repeat, these three
  cases are as follows. First when $A_{h}$ is negative, $m_B^2$
  positive and $B_{4, h}$ positive such that $D_h$ is positive. Second
  when $A_h$ is positive, $m_B^2$ is negative and $B_{4, h}$ negative
  such that $D_h$ is positive. Lastly when $A_h$ is positive and
  $m_B^2$ positive for either sign of $B_{4, h}$. }.

\subsection{unHiggsed branch} \label{uhp}

\subsubsection{Graphs of $U_{\rm eff}(\sigma_B)$ in various cases}

In this subsubsection we plot $U_{\rm eff}(\sigma_B)$ on the 
unHiggsed branch for various ranges of microscopic parameters.

We start with the following expression for the Landau-Ginzburg
potential in the unHiggsed branch in terms of the constants $A_u$ and
$B_{4,u}$:
\begin{equation}\label{fuh}
  U_{\rm eff}(\sigma_B) = \frac{N_B}{2 \pi} 
  \left( - A_u \frac{ 4 { \sigma}_B^3}{3} + 2 B_{4, u} { \sigma}_B^2  +  m_B^2 { \sigma}_B \right)\ , 
\end{equation}
where $A_u$ and $B_{4,u}$ are as in the first line of \eqref{abdef}.
Now recall that ${\sigma}_B$ is necessarily negative in the unHiggsed
phase. As a consequence \eqref{fuh} may be rewritten as
\begin{equation}\label{fuhn}
U_{\rm eff}(\sigma_B) = \frac{N_B}{2 \pi} 
\left(  A_u \frac{ 4 |{ \sigma}_B|^3}{3} + 2 B_{4, u} 
|{ \sigma}_B|^2 - m_B^2 |{\sigma}_B| \right) \ .
\end{equation}
The gap equation that follows by varying \eqref{fuhn} w.r.t.
${\sigma}_B$ is given in \eqref{exactgap} and is reproduced below:
\begin{equation}\label{hatcu}
{A_u} c_B^2 + 2 B_{4,u} c_B -m_B^2=0\ ,
\end{equation}
where $c_B=2 |{\sigma}_B|$. Note the formal and notational similarity
with the analogous equation \eqref{hatc} in the Higgsed phase. As in
the previous subsection we briefly analyse the behaviour of
\eqref{fuhn} as a function of ${\sigma}_B$ in all the various
cases. We define the discriminant $D_u$ of \eqref{hatcu}:
\begin{equation}\label{discuh}
D_u = 4(B_{4,u}^2 + m_B^2 A_u)\ .
\end{equation}
We then have the following cases as depicted in Figures \ref{unhigpos}
and \ref{unhigneg}:
\begin{enumerate}
\item \emph{$A_{u}$ is positive}: the potential $U_{\rm eff}$
  increases at large $|\sigma_B|$.
      \begin{figure}[htbp]
        \begin{center}
          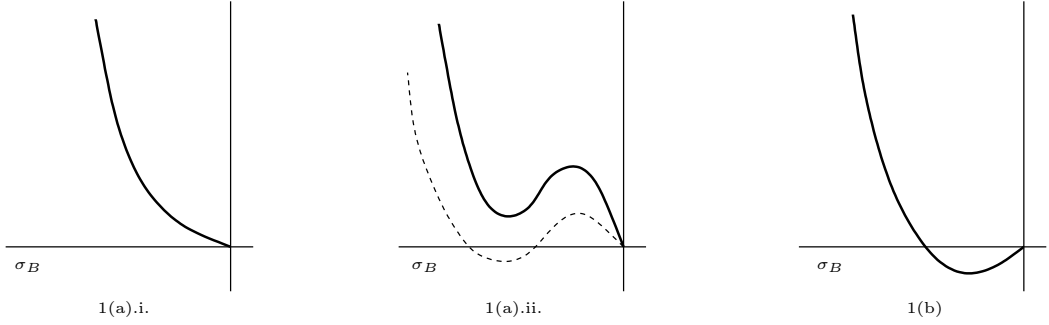
          \caption{Effective potential in the unHiggsed phase for $A_u$ positive.}
          \label{unhigpos}
        \end{center}
      \end{figure}
  \begin{enumerate}
  \item \emph{$m_B^2$ is negative}:
    \begin{enumerate}
    \item \emph{$B_{4,u}$ is positive, or $B_{4,u}$ is negative
        such that $D_u$ is negative}: the potential rises
      monotonically as $|\sigma_B|$ increases from zero to infinity,
      and there are no nontrivial positive solutions of the gap
      equation \eqref{hatcu}.
    \item \emph{$B_{4,u}$ is negative such that $D_u$ is
        positive}: As $|\sigma_B|$ is increased from zero,
      $U_{\rm eff}$ initially increases up to a local maximum and then
      decreases down to a local minimum and then increases without
      bound. The gap equation has two solutions the larger of which is
      the dominant stable phase (the smaller solution presumably
      describes unstable dynamics since it occurs at a local maximum
      of the effective potential).
    \end{enumerate}
  \item \emph{$m_B^2$ is positive}: For either sign of $B_{4,u}$,
    $U_{\rm eff}$ initially decreases, reaches a minimum and then
    turns and increases indefinitely. The gap equation has exactly one
    legal solution which is the dominant stable phase.
  \end{enumerate}
\item \emph{$A_{u}$ is negative}: the potential $U_{\rm eff}$
  decreases at large $|\sigma_B|$.
      \begin{figure}[htbp]
        \begin{center}
          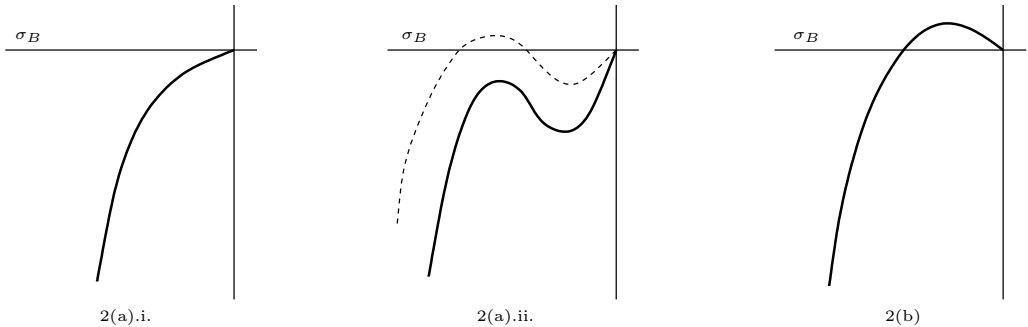
          \caption{Effective potential in the unHiggsed phase for $A_u$ negative.}
          \label{unhigneg}
        \end{center}
      \end{figure}
  \begin{enumerate}
  \item \emph{$m_B^2$ is positive}
    \begin{enumerate}
    \item \emph{$B_{4,u}$ is negative, or $B_{4,u}$ is positive
        such that $D_u$ is negative}: The potential decreases
      monotonically as $|\sigma_B|$ increases from zero to infinity,
      and there are no nontrivial positive solutions of the gap
      equation \eqref{hatcu}.
    \item \emph{$B_{4,u}$ is positive such that $D_u$ is
        positive}:  As $|\sigma_B|$ is increased from zero
      $U_{\rm eff}$ initially decreases down to a local minimum and
      then increases up to a local maximum after which it decreases
      without bound. The smaller of the two solutions to the gap
      equation \eqref{hatcu} is the `dominant' metastable phase (the
      larger solution presumably describes unstable dynamics again
      since it occurs at a local maximum of the potential).
    \end{enumerate}
  \item \emph{$m_B^2$ is negative}: For either sign of $B_{4,u}$,
    $U_{\rm eff}$ initially increases, reaches a maximum and then
    turns and decreases indefinitely. The gap equation has exactly one
    legal solution; this is a local maximum and so presumably
    describes an unstable `phase'.
  \end{enumerate}
\end{enumerate}
In the paragraphs above we have encountered three examples of
`unstable phases' (1.a.ii, 2.a.ii, 2.b). The solution of the gap
equation \eqref{hatcu} associated with each of these phases is easily
verified to be
\begin{equation}\label{roee}
c_B = \frac{-B_{4, u}-\sqrt{B_{4,u}^2+A_u m_B^2}}{A_{u}}\ .
\end{equation}
Moreover it is also easy to check that every legal (i.e. positive)
solution of the form \eqref{roee} is one of the three local maxima of
the paragraphs described above.

\subsubsection{Explanation for the instability of local maxima}

We have already explained above that
 \begin{equation}\label{signi} |{
    \sigma}_B|= -\frac{2\pi ({\bar \phi} \phi)_{\rm cl}}{N_B}\ ,
\end{equation}
where the RHS of this equation should be interpreted in quantum rather
than semiclassical terms since semiclassically the variable
$(\bar\phi \phi)_{\rm cl}$ is given by
$\langle \bar\phi\rangle\langle \phi\rangle$ and is positive. The
operator ${\bar \phi} \phi$ however is not necessarily positive (this
follows because the subtraction that is used to give this operator
meaning is not positive). Equation \eqref{signi} effectively asserts
that the unHiggsed branch explores only negative values of the
operator $\bar\phi \phi$. We have noted above the effective potential
as a function of $({\bar \phi} \phi)_{\rm cl}$ has unstable `phases'
that sit at local maxima of the effective potential. In the rest of
this subsection we will present an explanation of these instabilities.

Our proposal for the mechanism of the instability of the `local
maxima' phases is that it is the tachyonic instability of a bound
state of a single fundamental and antifundamental field in the singlet
channel. We claim that the solutions \eqref{roee} are all unstable in
this sense, while none of the stable phases - i.e. the phases that
occur at legal values of
\begin{equation}\label{roeen}
c_B = \frac{-B_{4, u}+\sqrt{B_{4,u}^2+A_u m_B^2}}{A_{u}}\ ,
\end{equation}
suffer from such an instability.

In order to see that this is indeed the case let us recall that bound
states do occur as poles in the S-matrix of a fundamental $\phi$ field
scattering off an antifundamental ${\bar \phi}$ field. Moreover these
poles do sometimes go tachyonic (i.e. their squared mass sometimes
goes below zero).  The condition for this to happen can be worked out
by following discussion in Section $4.5$ of \cite{Jain:2014nza} and
Appendix C of \cite{Yokoyama:2016sbx}.  The particle-antiparticle
scattering S-matrix has a pole with positive squared mass when
\begin{equation}
4\lambda_B \le \lambda_B^2 (3 x_6^B+4)-4\frac{b_4 \lambda_B}{c_B}\le 4\ .
\end{equation}
When 
$$  4\lambda_B = \lambda_B^2 (3 x_6^B+4)-4\frac{b_4 \lambda_B}{c_B}\ , $$
the pole is at threshold, i.e. $\sqrt{s}=2 c_B$.
On the other hand when 
\begin{equation}\label{poth}
 \lambda_B^2 (3 x_6^B+4)-4\frac{b_4 \lambda_B}{c_B}=4\ ,
\end{equation}
the pole lies at $\sqrt{s}=0$. Using the definitions \eqref{abdef} for
$A_u$ and $B_{4,u}$ it is easy to see that the condition \eqref{poth}
can be rewritten as
\begin{equation} \label{pothn}
B_{4,u}= - A_u c_B\ .
\end{equation}
(recall that the quantity $c_B$ is positive by definition). When
\begin{equation}\label{unpoth}
\lambda_B^2 (3 x_6^B+4)-4\frac{b_4 \lambda_B}{c_B}>4\ ,
\end{equation}
we have a bound state with negative squared mass, i.e.~a tachyonic
bound state. The condition for the existence of this tachyonic pole is
\begin{equation}\label{ctbs}
A_u c_B \leq -B_{4,u}\ .
\end{equation}
Of course the quantity $c_B$ is not independent of $A_u$ and 
$B_{4,u}$ but is determined in terms of these quantities by 
the gap equation. The solutions to the gap equation are given by  
\begin{equation}\label{roeege}
c_B = \frac{-B_{4, u} \pm \sqrt{B_{4,u}^2+A_u m_B^2}}{A_{u}}
\end{equation}
Inserting these solutions into the condition \eqref{ctbs}, we 
find that the condition \eqref{ctbs} is met whenever
\begin{equation}\label{roeege}
-B_{4, u} \pm \sqrt{B_{4,u}^2+A_u m_B^2} \leq -B_{4, u}
\end{equation}
This condition is obeyed by the `minus' branch of solutions
\eqref{roee} but not by the `plus' branch of solutions
\eqref{roeen}. But we have seen above that this is precisely the split
between the local maxima (solutions \eqref{roee}) and local minima
(solutions \eqref{roeen}) of the effective action \eqref{fuhn}. It is
thus natural to identify the tachyonic bound states as the explanation
for the instability of the `minus' branch of solutions \eqref{roee}.

It follows, in other words, that the instabilities in the unHiggsed
phase occur for the same reason as the instabilities in the Higgsed
phase, but for a different field. Unstable Higgsed `phases' occurred
when our solution to the gap equations was at a maximum of the
potential for the field $\phi$. We propose that instabilities in the
unHiggsed phase occur for solutions to the gap equation around maxima
for the field $(\bar\phi \phi)_{\rm cl}$ that is very ostensibly
related to the bound state of ${\bar \phi}$ and $\phi$.

In the case of the Higgsed theory in the $\lambda_B \to 0$ limit
\eqref{class} we obtained a classical theory (with an overall factor
of $\lambda_B^{-1}$ outside the action) in terms of the variable
$\varphi$. In the current unHiggsed context the effective potential
does not have a clear classical limit as $\lambda_B\to 0$. On
solutions to the gap equation that follows from varying \eqref{fuhn}
w.r.t.  $|{ \sigma}_B|$, it turns out that ${ \sigma}_B$ and
${\bar \phi} \phi$ (rather than $\lambda_B { \sigma}_B$ and
${\bar \varphi } \varphi$ as in the Higgsed phase) are finite as
$\lambda_B \to 0$. The field $\varphi$ which was the natural classical
variable at weak coupling in the Higgsed phase does not seem to be
useful in the analysis of the unHiggsed branch (this is probably a
reflection of the fact that dynamics is always quantum on this
branch).

\subsection[Landau-Ginzburg Analysis of the zero temperature phase
  diagram]{Landau-Ginzburg Analysis of the zero temperature phase
  diagram\protect\footnote{This subsection was 
  worked out in collaboration with O. Aharony.}}\label{phase}

In subsections \ref{hp} and \ref{uhp} we have already explored the
qualitative structure of the Landau-Ginzburg potential \eqref{LGpot},
plotted as a function of $\sigma_B$, separately for $\sigma_B>0$ and
$\sigma_B<0$. In this section we will simply put the analyses of
subsections \ref{hp} and \ref{uhp} together to obtain a global picture
of the Landau-Ginzburg potential as a function of $\sigma_B$ over all
possible ranges of parameters $x_6$, $\lambda_B b_4$ and $m_B^2$. We
reproduce the potential below. Recall that our exact Landau-Ginzburg
potential as a function of $\sigma_B$ (or equivalently, using
\eqref{frdnn}, a function of $(\bar\phi\phi)_{\rm cl}$) is given by
\begin{align}\label{LGpot1}
&U_{\rm eff}(\sigma_B) = \left\{\arraycolsep=1.4pt\def\arraystretch{2.2}\begin{array}{cl} \frac{N_B}{2 \pi} 
  \left[ (x_6 - \phi_2) \lambda_B^2 { \sigma}_B^3 + 2 \lambda_B b_4 { \sigma}_B^2  +  m_B^2 { \sigma}_B \right] & \quad\text{for }\sigma_B < 0\ , \\ \frac{N_B}{2\pi}\left[( x_6 - \phi_1) \lambda_B^2 { \sigma}_B^3 +  2 \lambda_B{ b}_4  { \sigma}_B^2 + { m}_B^2 { \sigma}_B\right]  & \quad\text{for }\sigma_B > 0\ . \end{array}\right.
\end{align}
Recall from \eqref{phi12def} that $\phi_1 < \phi_2$. As we have
explored in detail above, the plots of the effective potential are
qualitatively different when $x_6<\phi_1$, $\phi_1 <x_6< \phi_2$ and
$x_6 > \phi_2$. For this reason we analyse these three ranges of
parameters separately. It is also useful to recall the formulae for
the discriminants \eqref{disch} and \eqref{discuh} of the gap
equations \eqref{exactgap} in either phase of the theory:
\begin{align}
D_h &= \frac{4(2-|\lambda_B|)^2}{\lambda_B^2} \left[ (\lambda_B b_4)^2 - \tfrac{3 \lambda_B^2}{4}(x_6 - \phi_1)m_B^2\right]\ ,\nonumber\\
 D_u &= 4\left[(\lambda_B b_4)^2 - \tfrac{3\lambda_B^2}{4}(x_6 - \phi_2)  m_B^2\right]\ .
\end{align}

\subsubsection{Case I: $x_6> \phi_2$} \label{sso} In this case the
coefficient of $\sigma_B^3$ in the effective potential \eqref{LGpot1}
is positive both when $\sigma_B>0$ and when $\sigma_B<0$. It follows
that $U_{\rm eff}(\sigma_B)$ is an increasing function in the limits
$\sigma_B \to \pm \infty$. Note, in particular, that
$U_{\rm eff}(\sigma_B)$ is unbounded from below at large negative
$\sigma_B$ presumably indicating a runaway instability of the
theory. In other words, the theory has no truly stable phase in this
range of $x_6$. In this subsection we will sketch the `phase diagram'
of the theory, defined as the diagram that tracks the dominant
metastable phase as a function of the relevant parameters. \footnote{We
  emphasise that this phase diagram is formal; no phase - not even the
  dominant one - is stable. The theory always has a run away
  instability to tunnel to large negative values of $\sigma_B$.}  In
order to do this we simply plot $U_{\rm eff}(\sigma_B)$ as a function
of $\sigma_B$.  The detailed behaviour of the curve
$U_{\rm eff}(\sigma_B)$ at finite values of $\sigma_B$ depends on the
signs and values of $m_B^2$ and $\lambda_B b_4$.  We have the
following sub cases.

\begin{figure}[htbp]
  \begin{center}
    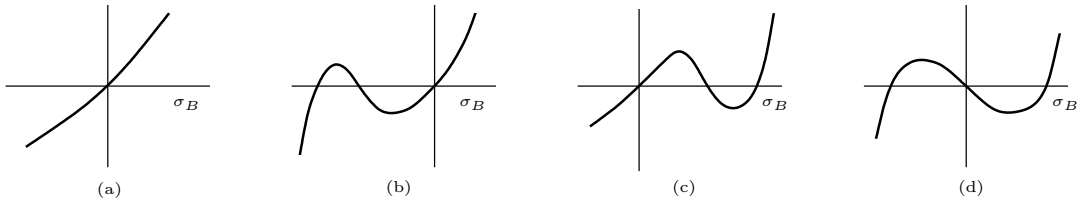
    \caption{Effective potential for $x_6 > \phi_2$.}
    \label{Case I}
  \end{center}
\end{figure}

\begin{enumerate}
\item  \emph{$m_B^2$ positive.}
  \begin{enumerate}
  \item \emph{$\lambda_B b_4$ positive with $D_u$ negative} or
    \emph{$\lambda_B b_4$ negative with $D_h$ negative}: In this
    case $U_{\rm eff}(\sigma_B)$ is a monotonically increasing
    function of $\sigma_B$ as depicted in Fig \ref{Case I}(a).
    $U_{\rm eff}(\sigma_B)$ has no extrema and so the gap equation has
    no solutions.
  \item \emph{$\lambda_B b_4$ positive with $D_u$ positive}: In
    this case the curve of $U_{\rm eff}(\sigma_B)$ takes the schematic
    form depicted in  Fig \ref{Case I}(b).
    $U_{\rm eff}(\sigma_B)$ has two extrema; a local minimum and a
    local maximum both for $\sigma_B<0$, so both in the unHiggsed
    branch. The local minimum is the only metastable phase of the
    theory (the maximum is unstable) and so is the dominant `phase'.
  \item \emph{$\lambda_B b_4$ negative with $D_h$ positive:} The
    graph of $U_{\rm eff}(\sigma_B)$ takes the schematic form depicted
    in Fig \ref{Case I}(c).  $U_{\rm eff}(\sigma_B)$ has two extrema;
    a local minimum and a local maximum both for positive $\sigma_B$
    so in the Higgsed branch. The local minimum is the only metastable
    phase of the theory (the maximum is unstable) and so is the
    dominant `phase'.
\end{enumerate}
\item \emph{$m_B^2$ negative, $\lambda_B b_4$ arbitrary:} In this
  case the graph of $U_{\rm eff}(\sigma_B)$ versus $\sigma_B$ takes
  the schematic form depicted in Fig \ref{Case I}(d). We
  have a local maximum at negative $\sigma_B$ (so in the unHiggsed
  phase) and a local minimum - so a metastable phase - at positive
  $\sigma_B$, so in the Higgsed branch. This local minimum is the
  dominant (metastable) phase.
\end{enumerate}

\begin{figure}[htbp]
  \begin{center}
    \resizebox{0.24\linewidth}{!}{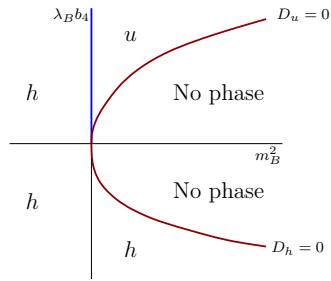}
    \caption{Phase diagram for $x_6 > \phi_2$. The positive
      $\lambda_B b_4$ axis (shown in blue) corresponds to a
      second-order phase transition.}
    \label{PhaseI}
  \end{center}
\end{figure}

Putting all this together we conclude that our theory has the 
(metastable) phase structure depicted in Fig. 8 of \cite{abcd} 
and redrawn here for convenience in Fig. \ref{PhaseI} of this paper.
Notice that whenever a metastable phase exists, a subdominant unstable
local maximum of $U_{\rm eff}(\sigma_B)$ also exists in the
vicinity. These are the subdominant `phases' that appear in Fig. 22(a)
of \cite{abcd}.

To end this subsubsection let us study what happens in the limit in
which $x_6 \to \phi_2$ from above. First, nothing special happens to
$U_{\rm eff}(\sigma_B)$ for positive $\sigma_B$. At negative
$\sigma_B$, however, the coefficient of the $\sigma_B^3$ term tends to
zero when $\sigma_B<0$. In this limit $D_u$ is always positive, so the
top half of the red curve in Fig.~\ref{PhaseI} tends to a horizontal
line (the $m_B^2 > 0$ axis).  Moreover, when $m_B^2$ and
$\lambda_B b_4$ are both positive (i.e. the case of Fig. \ref{Case
  I}(b)) the local minimum (which can be thought of as arising due to
a competition between the linear and quadratic terms in the action)
continues to occur at a fixed value of $\sigma_B$ and the
$U_{\rm eff}(\sigma_B)$ evaluated at this minimum also remains
fixed. But the local maximum of this diagram (which is a result of the
competition between the cubic and quadratic terms in the action) now
occurs at a value of $\sigma_B$ that tends to $-\infty$. Moreover the
value of $U_{\rm eff}(\sigma_B)$ at this maximum also tends to
$\infty$.  For $x_6\leq \phi_2$ this local maximum simply does not
exist any more.

\subsubsection{Case II: $\phi_1 < x_6 < \phi_2$} \label{sstw} In this
case, the coefficient of $\sigma_B^3$ is positive for $\sigma_B > 0$
and negative for $\sigma_B < 0$ which implies that the potential is
bounded below for all values of $\sigma_B$, so the theory is stable.
$U_{\rm eff}(\sigma_B)$ is a decreasing function of $\sigma_B$ for
large negative $\sigma_B$, but is an increasing function of $\sigma_B$
for large positive $\sigma_B$.
\begin{figure}[htbp]
  \begin{center}
    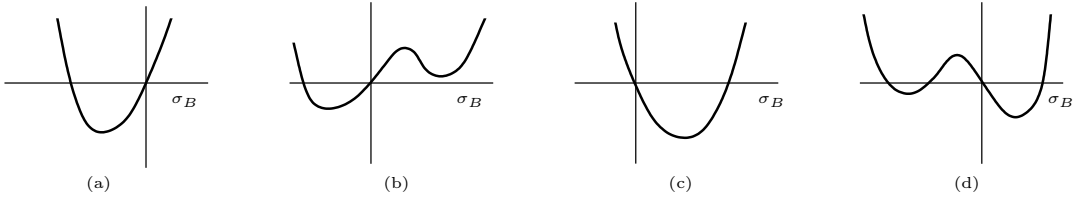
    \caption{Effective potential for $\phi_1 < x_6 < \phi_2$.}
    \label{Case II}
  \end{center}
\end{figure}
\begin{enumerate}
\item\emph{$m_B^2$ positive}: \emph{$\lambda_B b_4$ positive or
    $\lambda_B b_4$ negative with $D_h$ negative:} In this case the
  graph of $U_{\rm eff}(\sigma_B)$ versus $\sigma_B$ takes the form
  depicted in Fig \ref{Case II}(a). The global minimum in the
  unHiggsed phase (negative $\sigma_B$) is the only extremum of
  $U_{\rm eff}(\sigma_B)$; this phase dominates the phase diagram.

\item \emph{$m_B^2$ positive and $\lambda_B b_4$ negative with
    $D_h$ positive} or \\ \emph{$m_B^2$ negative and
    $\lambda_B b_4$ negative with $D_u$ positive}: In this case the
  graph of $U_{\rm eff}(\sigma_B)$ versus $\sigma_B$ takes the form
  depicted in Fig \ref{Case II}(b) when $m_B^2$ is positive and of the
  form depicted in Fig \ref{Case II}(d) when $m_B^2$ is negative. In
  either case the graph has a local minimum in the unHiggsed branch
  (negative $\sigma_B$) and a local minimum in the Higgsed branch
  (positive $\sigma_B$) separated by a local maximum. The maximum
  occurs in the Higgsed branch when $m_B^2>0$ but in the unHiggsed
  branch when $m_B^2<0$.  \footnote{In fact at $m_B^2=0$ the local
    maximum goes through $\sigma_B=0$; this maximum undergoes a
    `second order phase transition' at this point from the Higgsed to
    the unHiggsed phase. This point is depicted in Figure
    \ref{InterII}.}  The dominant phase is the local minimum with the
  smaller free energy. Which phase dominates depends on the precise
  values of $m_B^2$, $\lambda_B b_4$ and $x_6$. A detailed analysis
  has been performed in \cite{abcd} and we summarise the results
  here. When $x_6$ is strictly between $\phi_1$ and $\phi_2$ the
  theory has a first order phase transition line along the curve
  \begin{equation}
    D_\nu = m_B^2 - \nu_c(x_6) {(\lambda_B b_{4})^2} = 0\ .
  \end{equation}
  The function $\nu_c(x_6)$ was studied in detail in \cite{abcd}; see
  around Fig. 25 and Fig. 26. The function $\nu_c(x_6)$ is monotonically
  decreasing as a function of $x_6$ with $x_6 \in (\phi_1 , \phi_2)$.
  The function $\nu_c(x_6)$ is negative when $x_6$ is near $\phi_2$
  and hence the first order transition line is in the third quadrant
  (corresponding to Fig. \ref{PhaseII}(a)). When $x_6$ is near
  $\phi_1$, the function $\nu_c(x_6)$ is positive and hence the first
  order transition line is in the fourth quadrant
  (Fig. \ref{PhaseII}(b)). The phase transition line crosses over to
  the fourth quadrant from the third quadrant (equivalently,
  $\nu_c(x_6)$ goes from being negative to positive) at some
  intermediate value of $x_6$. This intermediate value occurs at
  $x_6 = \tfrac{1}{2}(\phi_1 + \phi_2)$ and the phase transition line
  coincides with the negative $\lambda_B b_4$ axis. We plot the phase
  diagram for this case and also the corresponding Landau-Ginzburg
  potential on the phase transition line in Figure
  \ref{InterII}. Clearly, we have two exactly equal minima and hence
  the onset of a first order phase transition.

\begin{figure}[htbp]
  \begin{center}
    \resizebox{0.8\linewidth}{!}{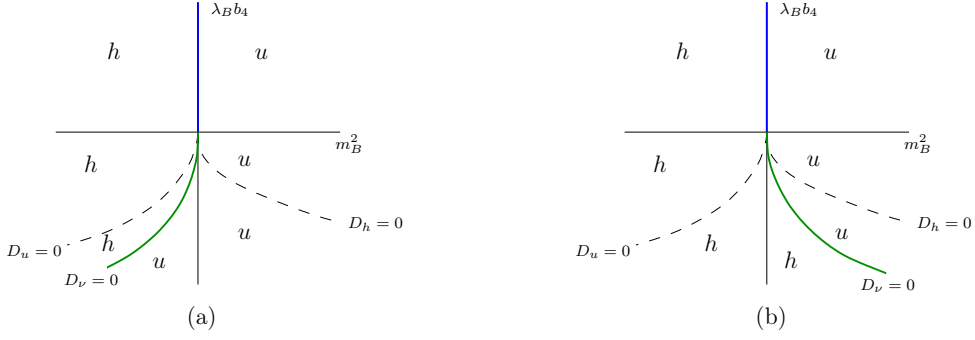}
    \caption{The phase diagram for $\phi_1 < x_6 < \phi_2$. There is a
      second order phase transition (shown in blue) along the positive
      $\lambda_B b_4$ axis. The first order phase transition line is
      the curve (shown in green) between the two dashed curves. The
      precise location of this phase transition curve varies as we
      change $x_6$. Two possible locations of this curve have been
      sketched in the two figures above. The first figure corresponds
      to $x_6$ near $\phi_2$ and the second figure corresponds to
      $x_6$ near $\phi_1$.}
    \label{PhaseII}
  \end{center}
\end{figure}

  \begin{figure}[htbp]
    \begin{center}
      \resizebox{0.8\linewidth}{!}{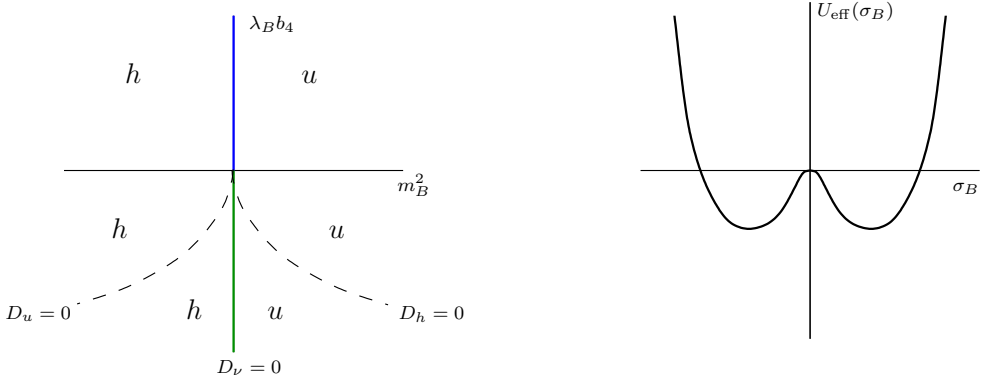}
      \caption{The first figure is the phase diagram for
        $x_6 = \tfrac{1}{2}(\phi_1 + \phi_2)$. The second figure is
        the Landau-Ginzburg potential at the same value of $x_6$ for a
        point on the first order phase transition line (green)
        corresponding to $m_B^2 = 0$ and some $\lambda_B b_4 < 0$.}
      \label{InterII}
    \end{center}
  \end{figure}

  Let us study the behaviour in the limits $x_6 \to \phi_1$ and
  $x_6 \to \phi_2$. In the limit $x_6 \to \phi_2$ from below, the
  unHiggsed branch minimum occurs at $\sigma_B \rightarrow -\infty$
  and the potential $U_{\rm eff}(\sigma_B)$ evaluated on this solution
  tends to $-\infty$. In this limit the unHiggsed branch local minimum
  is the dominant phase for every value of $\lambda_Bb_{4} < 0$ and
  $m_B^2$.  In the opposite limit $x_6 \to \phi_1$ the Higgsed branch
  local minimum occurs at very large values of $\sigma_B$ and
  $U_{\rm eff}(\sigma_B)$ evaluated on this solution tends to
  $-\infty$. In this limit the Higgsed branch local minimum is the
  dominant phase for every value of $\lambda_B b_{4} < 0$ and $m_B^2$.
	 
\item\emph{$m_B^2$ negative:} \emph{$\lambda_B b_4$ positive
    or $\lambda_B b_4$ negative with $D_u$ negative:} The graph of
  $U_{\rm eff}(\sigma_B)$ versus $\sigma_B$ takes the form depicted in
  Fig. \ref{Case II}(c). The global minimum in the Higgsed branch
  (positive $\sigma_B$) is the only extremum of
  $U_{\rm eff}(\sigma_B)$; this phase dominates the phase diagram.
\end{enumerate}
Putting all this together we arrive at the phase diagram 
presented in Fig. 7 of \cite{abcd}. This phase diagram is 
resketched in Fig \ref{PhaseII} for convenience.

\subsubsection{Case III: $x_6 < \phi_1$} \label{ssth} In this case,
the coefficient of $\sigma_B^3$ is negative for both $\sigma_B > 0$
and $\sigma_B < 0$. It follows that $U_{\rm eff}(\sigma_B)$ is a
increasing function in the limits $\sigma_B \to \pm \infty$. Note, in
particular, that $U_{\rm eff}(\sigma_B)$ is unbounded from below at
large positive $\sigma_B$ presumably implying a runaway instability of
the theory. In this case the instability is easy to understand as it
is present even in the classical theory at sufficiently negative
values of $x_6$.  Just as in Section \ref{sso}, in range of parameters
the RB the theory has no truly stable phases. As in Section \ref{sso},
in this subsubsection we will sketch the `phase diagram' of the theory,
defined as the diagram that tracks the dominant metastable phase
as a function of relevant parameters. As in Section \ref{sso} we read
off our results from plots of $U_{\rm eff}(\sigma_B)$ as a function of
$\sigma_B$. We have the following subcases.

\begin{figure}[htbp]
	\begin{center}
		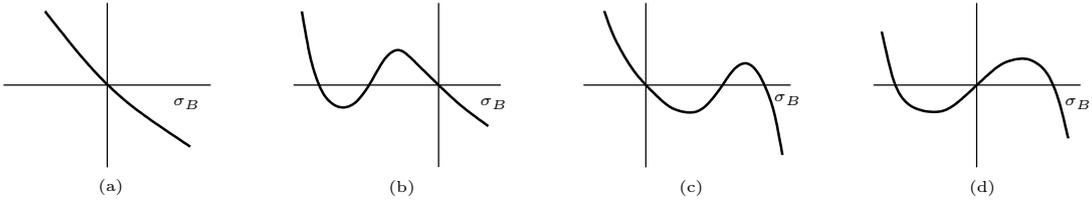
		\caption{Effective potential for $x_6 < \phi_1$.}
		\label{Case 3}
	\end{center}
\end{figure}

\begin{enumerate}
\item \emph{$m_B^2$ negative.}
  \begin{enumerate}
  \item \emph{ $\lambda_B b_4$ negative with $D_u$ negative or
      $\lambda_B b_4$ positive with $D_h$ negative:}
    $U_{\rm eff}(\sigma_B)$ is a monotonically decreasing function of
    $\sigma_B$ as depicted in  Fig \ref{Case
      3}(a). $U_{\rm eff}(\sigma_B)$ has no extrema, and so the gap
    equation has no solutions.
  \item \emph{$\lambda_B b_4$ negative with $D_u$ positive:} In
    this case the curve of $U_{\rm eff}(\sigma_B)$ takes the schematic
    form depicted in Fig. \ref{Case 3}(b).
    $U_{\rm eff}(\sigma_B)$ has two extrema; a local minimum and a
    local maximum both for $\sigma_B<0$, so both in the unHiggsed
    phase. The local minimum is the only metastable phase of the
    theory (the maximum is unstable) and so is the dominant `phase'.
  \item \emph{ $\lambda_B b_4$ positive with $D_h$ positive:} The
    graph of $U_{\rm eff}(\sigma_B)$ takes the schematic form depicted
    in Fig \ref{Case 3}(c).  $U_{\rm eff}(\sigma_B)$
    has two extrema; a local minimum and a local maximum both for
    positive $\sigma_B$ so in the Higgsed phase. The local minimum is
    the only metastable phase of the theory (the maximum is unstable)
    and so is the dominant `phase'.
  \end{enumerate}
\item \emph{$m_B^2$ positive, $\lambda_B b_4$ arbitrary:} In this
  case the graph of $U_{\rm eff}(\sigma_B)$ versus $\sigma_B$ takes
  the form depicted in Fig \ref{Case 3}(d). We have a
  local minimum in at negative $\sigma_B$ (so in the unHiggsed phase)
  and a local maximum at positive $\sigma_B$, so in the Higgsed
  phase. This local minimum is the dominant (metastable) phase.
\end{enumerate}

\begin{figure}[htbp]
	\begin{center}
		\resizebox{0.25\linewidth}{!}{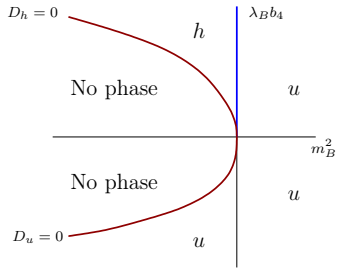}
		\caption{Phase diagram for $x_6 < \phi_1$.  The
                  positive $\lambda_B b_4$ axis (shown in blue)
                  corresponds to a second-order phase transition.}
		\label{Phase3}
	\end{center}
\end{figure}

Putting all this together we conclude that our theory has the 
(metastable) phase structure depicted in Fig. 9 of \cite{abcd} 
and redrawn here for convenience in Fig. \ref{Phase3} of this paper.

Notice that whenever a metastable phase exists, a subdominant unstable
local maximum of $U_{\rm eff}(\sigma_B)$ also exists in the
vicinity. These are the subdominant `phases' that appear in Fig. 27(b)
of \cite{abcd}.

To end this subsubsection let us study what happens in the limit in
which $x_6 \to \phi_1$ from above. For negative values of $\sigma_B$,
nothing special happens to $U_{\rm eff}(\sigma_B)$. At positive
$\sigma_B$, however, the coefficient of the $\sigma_B^3$ term tends to
zero. In this limit $D_h$ is always positive, so the top half of the
red curve in Fig.  \ref{Phase3} tends to a horizontal line (the
negative $m_B^2$ axis).  Moreover, when $m_B^2$ and $\lambda_B b_4$
are both positive (i.e. the case of sub Fig. \ref{Case 3}(d)) the
local minimum (which can be thought of as arising due to a competition
between the linear and quadratic terms in the action) continues to
occur at a fixed value of $\sigma_B$ and the $U_{\rm eff}(\sigma_B)$
evaluated at this minimum also remains fixed. But the local maximum of
this diagram (which is a result of the competition between the cubic
and quadratic terms in the action) now occurs at a value of $\sigma_B$
that tends to $\infty$. Moreover the value of $U_{\rm eff}(\sigma_B)$
at this maximum also tends to $\infty$.  For $x_6 \geq \phi_1$ this
local maximum simply does not exist any more.

\section{Discussion} 

The results of this paper suggest several questions for future
work. First, it would be interesting to generalise the computation of
S-matrices presented in \cite{Jain:2014nza, Yokoyama:2016sbx} to the
Higgsed phase of the Regular Boson theory. The fact that this (and
related) computations may throw light on the dual fermionic
interpretation of the $Z$ boson - as discussed in detail in
\cite{Choudhury:2018iwf} - make it particularly interesting.

One of the most interesting results of this paper is the off-shell
effective action \eqref{LGpot}. It would be interesting to generalise
this result to finite values of temperature and chemical potential and
explicitly observe the smoothing-out of the non-analyticity which was
present at zero temperature. It would be also interesting to compute a
similar action for the theory of one fundamental boson and one
fundamental fermion studied in \cite{Jain:2013gza} and to use this
action to unravel the phase structure of the deformed ${\cal N}=2$
supersymmetric matter Chern-Simons theory with a single chiral
multiplet in the fundamental representation.  It is possible that such
an investigation will have interesting interplays with supersymmetry:
for example it may be possible to find a superspace version of
\eqref{LGpot}.

In Section \ref{ose} we have presented a three-variable off-shell free
energy that reproduces the gap equation and thermal free energy of the
regular boson theory. We have also presented a physical interpretation
of the variable $\sigma_B$ that enters this action. It would be
interesting to investigate whether there are interesting off-shell
interpretations of the other dynamical variables - $c_B$ and
${\tilde \cS}$ - that appear in Section \ref{ose}.

Above, we have found a preferred value for the cosmological constant
counterterm $\Lambda$ of the CB theory - one that correctly reproduces
the tadpole condition for the regular boson theory (see
\eqref{alphapred}). It would be interesting to derive
\eqref{alphapred} from a more fundamental physical principle. It would
also be interesting to investigate if this result makes any physical
predictions for the critical boson theory: is it correct, for
instance, to interpret the Legendre transform of \eqref{osfeeh} w.r.t
$m_B^{\rm cri}$ (with $\Lambda$ chosen to have the value
\eqref{alphapred}) as the Coleman-Weinberg potential of the CB theory
w.r.t its dimension two scalar operator $J_0$? It would be interesting
to further investigate this and similar questions, and their
implications.

Finally it would be interesting to generalise the considerations of
this paper - even qualitatively - to finite values of $N_B$. We leave
all these questions for future work.

\acknowledgments

We would like to thank F.~Benini and D.~Radicevic for useful
discussions. We would especially like to thank O. Aharony for
collaboration on section \ref{phase}, for several very useful
discussions over the course of many months and for very useful
comments on a preliminary version of this manuscript.  The work of
A.~D., I.~H., L.~J., S.~M., and N.~P. was supported by the Infosys
Endowment for the study of the Quantum Structure of Spacetime.
S.~J. would like to thank TIFR, Mumbai for hospitality during the
completion of the work. The work of S.~J. is supported by the
Ramanujan Fellowship.  Finally we would all like to acknowledge our
debt to the steady support of the people of India for research in the
basic sciences.

\appendix

\section{Previously known results for the large $N$ free
  energy} \label{review} 
Below we will encounter several equations that involve the quantities
\begin{equation}\begin{split}
    \cC =&\frac{1}{2} \int_{-\pi}^\pi d\alpha\,\rho_F(\alpha)
    \( \log(2 \cosh (\tfrac{{\hat c}
      _F +i \alpha }{2}))+ \log(2 \cosh (\tfrac{{\hat c}_F - i\alpha }{2})) \),\\
    \cS =&\frac{1}{2}\int_{-\pi}^\pi d\alpha\,\rho_B(\alpha)
    \( \log(2 \sinh (\tfrac{{\hat c}_B +i\alpha }{2}))+ \log(2 \sinh
    (\tfrac{{\hat c}_B - i\alpha }{2})) \),
\label{nss}
\end{split}\end{equation}
where $\hat{c}_B$ and $\hat{c}_F$ are (dimensionless versions of) the
thermal masses in the boson and fermion theory respectively.

Using \eqref{nred}, it is not difficult to verify the following
identities:
\begin{equation}\label{ndualofq1}
\lambda_B \cS=-\frac{\sgn(\lambda_F)}{2}c_F+\lambda_F \cC\ ,\quad
\lambda_F \cC=-\frac{\sgn(\lambda_B)}{2}c_B+\lambda_B \cS\ .
\end{equation}

\subsection{Results for the critical fermion theory}
In \cite{Minwalla:2015sca} the `fixed holonomy'
${\mathbb R}^2 \times S^1$ partition function - $v_F[\rho_F]$ - of the
fermionic theory has been evaluated in both fermionic phases. The
final result of this calculation is most conveniently given in terms
of an auxiliary off-shell free energy
\begin{align}\label{ofcrf12}
&F_F(c_F, \zeta_F, \tl\cC)\nonumber\\
 &=\frac{N_F }{6\pi} \Bigg[ {\hat c}_F^3-2\lambda_F^2 {\tilde \cC}^3 -\frac{3}{2} \left({\hat c}_F^2 -\frac{16\pi^2}{\kappa_F^2}{\hat \zeta}_F^2 \right) {\tilde \cC} +\frac{6\pi {\hat y}_2^2}{\kappa_F\lambda_F}{\hat \zeta}_F-\frac{24 \pi^2 {\hat y}_4}{\kappa_F^2\lambda_F}{\hat \zeta}_F^2+\frac{24 \pi^3 x^F_6}{\kappa_F^3\lambda_F}{\hat \zeta}_F^3 \nonumber\\
&\qquad\qquad - \frac{3}{2} {\tilde \cC} \left( {\hat c}_F^2 -\left({2\lambda_F} {\tilde \cC} -\frac{4\pi}{\kappa_F}\hat\zeta_F\right)^2  \right)\nonumber \\
&\qquad\qquad-3 \int_{-\pi}^{\pi}d\alpha \rho_{F}(\alpha)\int_{{\hat c}_F}^{\infty} dy ~y~\left( \ln\left(1+e^{-y-i\alpha}\right)  + \ln\left(1+e^{-y+i\alpha}\right)\right)\Bigg].
\end{align}
The auxiliary off-shell free energy \eqref{ofcrf12} is a function of
three variables - $c_F$, ${\zeta}_F$ and ${\tilde \cC}$ - in addition
to the temperature and the holonomies. The free energy $v_F[\rho_F]$
defined in \eqref{nseff} is obtained from $F_F[\rho_F]$ in
\eqref{ofcrf12} by extremizing the latter quantity w.r.t.~these three
`dynamical' variables. Extremizing the free energy \eqref{ofcrf12}
w.r.t.~the variable ${\tilde \cC}$ yields the equation of motion \be
{\hat c}_F^2 =\left({2\lambda_F} {\tilde \cC}
  -\frac{4\pi}{\kappa_F}\hat\zeta_F\right)^2\ .\label{cfng} \ee Varying
w.r.t. $c_F$ yields
\begin{equation}
\label{vcf} 
{\tilde \cC}= \cC\ ,
\end{equation}
with $\cC$ given in \eqref{nss}, while the stationarity of variation
w.r.t $\zeta_F$ yields
\begin{equation}
  \label{sigsol}-\frac{3}{4} \left(\frac{4\pi \hat\zeta_F}{\kappa_F}\right)^2
   x_6^F-\frac{16\pi\hat\zeta_F}{\kappa_F}  \lambda_F{ {\tilde \cC}}+
  \frac{8\pi\hat\zeta_F}{\kappa_F}  \hat y_4 + 4 \lambda_F^2{ {\tilde \cC}}^2-\hat
  y_2^2=0\ .
\end{equation}
Plugging \eqref{vcf} into \eqref{cfng} and \eqref{sigsol} respectively
yields the simplified gap equations
\begin{equation}
{\hat c}_F^2 = \left({2\lambda_F} { \cC} -\frac{4\pi \hat\zeta_F}{\kappa_F}\right)^2\ ,
\label{1cf}
\end{equation}
and 
\begin{equation}
  \label{sigsoln}-\frac{3}{4} \left(\frac{4\pi \hat\zeta_F}{\kappa_F}\right)^2
   x_6^F-\frac{16\pi\hat\zeta_F}{\kappa_F}  \lambda_F{ {\cC}}+
  \frac{8\pi \hat\zeta_F}{\kappa_F}  \hat y_4 + 4 \lambda_F^2{ {\cC}}^2-\hat
  y_2^2=0\ ,
\end{equation}
for the quantities $c_F$ and $\zeta_F$.

Below we will find it useful to work with a reduced off-shell free energy, obtained by integrating ${\tilde \cC}$ out of
\eqref{ofcrf12}. In order to do this we note that \eqref{cfng} has two
solutions
\begin{equation}\label{cCto}
2 \lambda_F {\tilde \cC}= \frac{4 \pi {\hat \zeta}_F}{\kappa_F} \pm{\hat c}_F\ ,
\end{equation}
The undetermined sign in \eqref{cCto} is completely free.  
Clearly this sign is (tautologically)  given by 
${\rm sgn}({\tilde X}_F)$
where
\begin{equation}
\tl{X}_F \equiv {2\lambda_F} {\tilde \cC} -\frac{4\pi\hat\zeta_F}{\kappa_F}\ .
\end{equation}
It follows that \eqref{cCto} may formally be rewritten as
\begin{equation}\label{cCt}
2 \lambda_F {\tilde \cC}= \frac{4 \pi {\hat \zeta}_F}{\kappa_F}+{\rm sgn}({\tilde X}_F){\hat c}_F\ .
\end{equation}
The reduced free energy - which is now a function only of two
variables $c_F$ and $\zeta_F$ - is obtained by plugging either of
these two solutions into \eqref{ofcrf12}. Note that when we do this
the second line of \eqref{ofcrf12} vanishes as a consequence of
\eqref{cfng}.\footnote{This reduced form of the off-shell free energy
  - rather than the fully off-shell free energy \eqref{ofcrf12} - was
  presented in \cite{Minwalla:2015sca}. The off-shell free energy
  \eqref{ofcrf12} is a new formula that has not previously been
  presented in the literature.} Inserting the solutions \eqref{cCt}
into the free energy \eqref{ofcrf12}, we have the following explicit
expression for the reduced off-shell free energy as a function of
$c_F$ and $\zeta_F$
\begin{align}\label{ofcrf}
  F_F(c_F, \zeta_F) 
&=\frac{N_F }{6\pi} \Bigg[ \frac{|\lambda_F| - \sgn(\lambda_F)\sgn(\tl{X}_F)}{|\lambda_F|} \hat{c}_F^3 - \frac{3}{2\lambda_F}\left(\frac{4\pi\hat\zeta_F}{\kappa_F} \right) (\hat{c}_F^2 - {\hat y}_2^2) \nonumber \\ 
&\qquad\qquad   -\frac{3 {\hat y}_4}{2\lambda_F}\left(\frac{4\pi {\hat \zeta}_F}{\kappa_F}\right)^2 +\frac{(3  x^F_6 + 4)}{8\lambda_F}\left(\frac{4\pi {\hat \zeta}_F}{\kappa_F}\right)^3\nonumber \\  
&\qquad\qquad - 3 \int_{-\pi}^{\pi}d\alpha\, \rho_{F}(\alpha)\int_{{\hat c}_F}^{\infty} dy ~y~\left( \ln\left(1+e^{-y-i\alpha}\right)  + \ln\left(1+e^{-y+i\alpha}\right)\right)\Bigg].
\end{align}
Note that the above reduced off-shell free energy function has two
branches depending on the sign ${\rm sgn}({\tilde X}_F)$.  We refer to
the branch in which ${\rm sgn}({\tilde X}_F) \sgn(\lambda_F) >0$ as
the unHiggsed branch, and the branch in which
${\rm sgn}({\tilde X}_F) \sgn(\lambda_F) < 0$ as the Higgsed branch.
It is easily verified that the variation of \eqref{ofcrf} with respect
to ${\hat c}_F$ yields \eqref{1cf} while the variation of
\eqref{ofcrf} w.r.t $\zeta_F$ yields \eqref{sigsoln}.

Note that that the gap equations \eqref{cfng}, \eqref{vcf}
(equivalently \eqref{1cf}) and \eqref{sigsol} - unlike the free energy
\eqref{ofcrf} - have no explicit dependence on
${\rm sgn}(\tilde X_F)$. Nonetheless the same gap equations hold for
both `phases' of the theory, i.e.~for both choices of
${\rm sgn}(\tilde X_F \lambda_F)$.\footnote{The fact that
  ${\rm sgn}(\tilde X_F)$ disappears from the gap equations is obvious
  when we obtain the equations from \eqref{ofcrf12}, even though this
  fact might appear mysterious when derived starting from
  \eqref{ofcrf}.}  It follows from this observation that the solutions
to the finite temperature gap equations of the fermionic theory vary
analytically as we pass from one `phase' to another. In fact more is
true; the finite temperature free energy of the fermionic theory is
itself analytic as one passes from the unHiggsed to the Higgsed
phase. At the physical level, the sharp zero temperature distinction
between the Higgsed and unHiggsed phases gets blurred out by finite
temperature effects.

\subsection{Results for regular bosons in the unHiggsed Phase}
The off-shell free energy for the RB theory was computed in
 \cite{Minwalla:2015sca} and is given by
\begin{equation}\label{noffb12}
\begin{split}
F_B(c_B, \tl\cS) &= \frac{N_B}{6\pi} \Bigg[ -{\hat c}_B^3+2\left({\hat c}_B^2-{\hat m}_B^2 \right){\tilde \cS}+2\lambda_B {\hat b_4} {\tilde \cS}^2 \\
&\qquad\qquad + {\tilde \cS} \left( {{\hat c}_B^2}-\hat m_B^2 - (4 + 3 x_6^B)\lambda_B^2 {\tilde \cS}^2 + 4\lambda_B \hat b_4 {\tilde \cS}   \right) 
\\
&\qquad\qquad+ 3 \int_{-\pi}^{\pi} d\alpha \rho_{B}(\alpha)\int_{{\hat c}_B}^{\infty} dy ~y~\left( \ln\left(1-e^{-y-i\alpha}\right)  + \ln\left(1-e^{-y+i\alpha}\right)\right)\Bigg],
\end{split}
\end{equation}
or equivalently by 
\begin{equation}\label{bosoff}
\begin{split}
F_B(c_B, \tl\cS) &= \frac{N_B}{6\pi} \Bigg[ -{\hat c}_B^3+ 3 {\tilde \cS} \left({\hat c}_B^2-{\hat m}_B^2 \right) + 6 {\hat b_4} \lambda_B  {\tilde \cS}^2 - (4+3 x_6^B)\lambda_B^2 {\tilde \cS}^3 \\
&\qquad\qquad+ 3 \int_{-\pi}^{\pi} d\alpha \rho_{B}(\alpha)\int_{{\hat c}_B}^{\infty} dy ~y~\left( \ln\left(1-e^{-y-i\alpha}\right)  + \ln\left(1-e^{-y+i\alpha}\right)\right)\Bigg]\ .
\end{split}
\end{equation}
As in the previous subsection, the free energy $v_B[\rho_B]$ defined in
\eqref{nseff} is obtained by extremizing the action \eqref{noffb12}
w.r.t. the dynamical variables ${\tilde \cS}$ and $c_B$. The equations
of motion that follow by varying \eqref{noffb12} w.r.t ${\tilde \cS}$
and $c_B$ respectively are
\begin{align}
{{\hat c}_B^2} 
&= (4 +3 x_6^B)\lambda_B^2 {\tilde \cS}^2 -4\lambda_B \hat b_4 {\tilde \cS}  +\hat m_B^2\ ,\label{scalargapequationo}
\end{align}
and 
\begin{equation}
\label{cbv}
{\tilde \cS}= \cS\ ,
\end{equation}
where $\cS$ was defined in \eqref{nss}.  Inserting \eqref{cbv} into
\eqref{scalargapequationo} yields the gap equation for the single
variable $c_B$
\begin{align}
  {{\hat c}_B^2} 
  &= (4 + 3 x_6^B)\lambda_B^2 { \cS}^2 -4\lambda_B \hat b_4 {\cS}  +\hat m_B^2\ .     \label{scalargapequationt}
\end{align}
 As in the previous subsection it is possible to obtain a reduced free
energy by integrating ${\tilde \cS}$ out of \eqref{noffb12}. This may
be achieved by using \eqref{scalargapequationo} to solve for
${\tilde \cS}$ as a function of $c_B$ and plugging this solution into
\eqref{noffb12}.  \footnote{Note that the second line of
  \eqref{noffb12} vanishes when we do this as a this line is
  proportional to \eqref{scalargapequationo}. } This is the form in
which the off-shell free energy for the scalar theory in the unHiggsed
phase was presented in \cite{Minwalla:2015sca}.

\subsection{Duality in the unHiggsed phase and a prediction for the Higgsed phase}
We first list the duality map between various quantities in the RB and
CF theories. Recall that the 't Hooft parameters $\lambda_B$ and
$\lambda_F$ were defined as
\begin{equation}
\lambda_B = \frac{N_B}{\kappa_B}\ ,\quad \lambda_F = \frac{N_F}{\kappa_F}\ .
\end{equation}
The duality maps different parameters as follows:
\begin{align}\label{dualitymap}
  & N_F = |\kappa_B| - N_B\ ,\quad \kappa_F = -\kappa_B\ ,\quad \lambda_F=\lambda_B-{\rm sgn}(\lambda_B)\ ,\nonumber\\
  &x_6^F =x_6^B\ , \quad y_4 = b_4\ , \quad y_2^2 = m_B^2\ ,\quad |\lambda_B| \rho_B (\alpha)+ |\lambda_F| \rho_F(\pi - \alpha) = \frac{1}{2 \pi}\ .
\end{align}
The last relation gives rise to
\begin{equation}\label{dualhol}
N_F \rho_F(\alpha)  = \frac{|\kappa_B|}{2\pi} - N_B \rho_B(\pi-\alpha)\ .
\end{equation}
Consider the off-shell free energies for the critical fermion theory
in terms of the two `fields' $c_F$ and $\zeta_F$ given in
\eqref{ofcrf}:
\begin{align}\label{feroff}
F_F(c_F, \zeta_F) &=\frac{N_F }{6\pi} \Bigg[ \frac{\lambda_F - \sgn(\tl{X}_F)}{\lambda_F} {\hat c}_F^3  - \frac{3 }{2\lambda_F} \left(\frac{4\pi\hat\zeta_F}{\kappa_F}\right) (\hat{c}_F^2 - \hat{y}_2^2) \nonumber \\ &\qquad\qquad -\frac{3 {\hat y}_4}{2\lambda_F}\left(\frac{4\pi{\hat \zeta}_F}{\kappa_F}\right)^2   +\frac{(3  x^F_6 + 4)}{8\lambda_F}\left(\frac{4\pi{\hat \zeta}_F}{\kappa_F}\right)^3\nonumber + \\  &\qquad\qquad - 3 \int_{-\pi}^{\pi}d\alpha \rho_{F}(\alpha)\int_{{\hat c}_B}^{\infty} dy ~y~\left( \ln\left(1+e^{-y-i\alpha}\right)  + \ln\left(1+e^{-y+i\alpha}\right)\right)\Bigg].
\end{align}
Using the relation \eqref{dualhol}, the last line in \eqref{feroff}
can be rewritten as \footnote{The two terms in the RHS of
  \eqref{dualhol} simplify as follows. The integral over $\alpha$ in
  the term proportional to $|\kappa_B|$ can be performed by
  Taylor-expanding the logarithms and gives zero since the integrals
  are of the form $\int_{-\pi}^\pi d\alpha\, e^{i n \alpha} = 0$ for
  non-zero integers $n$. In the term proportional to
  $\rho_B(\pi-\alpha)$, we have performed the variable change
  $\alpha \to \pi -\alpha$ resulting in an additional minus sign in
  the argument of the logarithms.}
\begin{multline}
-\frac{ 3 N_F}{6\pi} \int_{-\pi}^{\pi}d\alpha \rho_{F}(\alpha)\int_{{\hat c}_F}^{\infty} dy ~y~\left( \ln\left(1+e^{-y-i\alpha}\right)  + \ln\left(1+e^{-y+i\alpha}\right)\right) =\\
= \frac{ 3 N_B}{6\pi} \int_{-\pi}^{\pi}d\alpha \rho_{B}(\alpha)\int_{{\hat c}_F}^{\infty} dy ~y~\left( \ln\left(1-e^{-y-i\alpha}\right)  + \ln\left(1-e^{-y+i\alpha}\right)\right)\ .
\end{multline}
Substituting the various fermionic parameters with their bosonic
counterparts in \eqref{dualitymap}, we get the following expression
for the dual of the fermionic off-shell free energy in terms of two
`fields' $c_F$ and $\zeta_F$:
\begin{align}\label{ferdual}
\cF(c_F, \zeta_F) = &\frac{N_B }{6\pi} \Bigg[- \frac{\lambda_B - \sgn(\lambda_B) - \sgn(\tl{X}_F)}{\lambda_B} {\hat c}_F^3  + \frac{3 }{2\lambda_B} \left(\frac{4\pi\hat\zeta_F}{\kappa_F}\right) (\hat{c}_F^2 - \hat{m}_B^2)  + \nonumber \\ &\qquad\qquad +\frac{3 {\hat b}_4}{2\lambda_B}\left(\frac{4\pi{\hat \zeta}_F}{\kappa_F}\right)^2   -\frac{(3  x^B_6 + 4)}{8\lambda_B}\left(\frac{4\pi{\hat \zeta}_F}{\kappa_F}\right)^3\nonumber + \\  &\qquad\qquad + 3 \int_{-\pi}^{\pi}d\alpha \rho_{B}(\alpha)\int_{{\hat c}_F}^{\infty} dy ~y~\left( \ln\left(1-e^{-y-i\alpha}\right)  + \ln\left(1-e^{-y+i\alpha}\right)\right)\Bigg].
\end{align}

\subsubsection{unHiggsed phase: ${\rm sgn}(X_F)=-{\rm sgn}(\lambda_B)$} 
In this case $\cF(c_F, \zeta_F)$ in \eqref{ferdual} simplifies to
\begin{align}\label{ferdualn}
\cF(c_F, \zeta_F) & = \frac{N_B }{6\pi} \Bigg[- {\hat c}_F^3  + \frac{3 }{2\lambda_B} \left(\frac{4\pi\hat\zeta_F}{\kappa_F}\right) (\hat{c}_F^2 - \hat{m}_B^2)  + \nonumber \\ &\qquad\qquad +\frac{3 {\hat b}_4}{2\lambda_B}\left(\frac{4\pi{\hat \zeta}_F}{\kappa_F}\right)^2   -\frac{(3  x^B_6 + 4)}{8\lambda_B}\left(\frac{4\pi{\hat \zeta}_F}{\kappa_F}\right)^3\nonumber + \\  &\qquad\qquad + 3 \int_{-\pi}^{\pi}d\alpha \rho_{B}(\alpha)\int_{{\hat c}_F}^{\infty} dy ~y~\left( \ln\left(1-e^{-y-i\alpha}\right)  + \ln\left(1-e^{-y+i\alpha}\right)\right)\Bigg].
\end{align}
If we now perform the field redefinitions
\begin{equation}\label{fieldredef}
\hat{c}_F = \hat{c}_B\ ,\quad \frac{4 \pi {\hat \zeta_F}}{\kappa_F} = 2 \lambda_B \tilde\cS\ .
\end{equation} 
we see that the fermionic off-shell free energy reduces exactly to the
regular boson off-shell free energy \eqref{bosoff}, establishing the
duality of the CF and RB theories in their unHiggsed phases. The
matching of off-shell free energies between the two theories
automatically guarantees the matching of gap equations, as the latter
are obtained by extremizing the off-shell free energies w.r.t. their
`fields'.

\subsubsection{Higgsed phase: ${\rm sgn}(X_F)={\rm sgn}(\lambda_B)$} 
In this case $\cF(c_F, \zeta_F)$ simplifies to 
\begin{align}\label{ferdualnn}
  \cF(c_F, \zeta_F) &= \frac{N_B }{6\pi} \Bigg[- \frac{\lambda_B - 2 \sgn(\lambda_B)}{\lambda_B} {\hat c}_F^3  + \frac{3 }{2\lambda_B} \left(\frac{4\pi\hat\zeta_F}{\kappa_F}\right) (\hat{c}_F^2 - \hat{m}_B^2)  + \nonumber \\ &\qquad\qquad +\frac{3 {\hat b}_4}{2\lambda_B}\left(\frac{4\pi{\hat \zeta}_F}{\kappa_F}\right)^2   -\frac{(3  x^B_6 + 4)}{8\lambda_B}\left(\frac{4\pi{\hat \zeta}_F}{\kappa_F}\right)^3\nonumber + \\  &\qquad\qquad + 3 \int_{-\pi}^{\pi}d\alpha \rho_{B}(\alpha)\int_{{\hat c}_F}^{\infty} dy ~y~\left( \ln\left(1-e^{-y-i\alpha}\right)  + \ln\left(1-e^{-y+i\alpha}\right)\right)\Bigg].
\end{align}
If we now make the field redefinitions
\begin{equation}
  \hat{c}_F = \hat{c}_B\ ,\quad \frac{4\pi\hat\zeta_F}{\kappa_F} = -2 \lambda_B {\hat \sigma}_B\ ,
\end{equation}
we find that \eqref{ferdualnn} reduces to
\begin{align}\label{ferdualm}
 F_B(c_B, \sigma_B) & = \frac{N_B }{6\pi} \Bigg[- \frac{\lambda_B - 2\sgn(\lambda_B)}{\lambda_B} {\hat c}_B^3  - 3 {\hat \sigma}_B (\hat{c}_F^2 - \hat{m}_B^2) + 6{\hat b}_4 \lambda_B{\hat\sigma}_B^2  + (3  x^B_6 + 4)\lambda_B^2{\hat\sigma}_B^3\nonumber \\  &\qquad\qquad + 3 \int_{-\pi}^{\pi}d\alpha \rho_{B}(\alpha)\int_{{\hat c}_B}^{\infty} dy ~y~\left( \ln\left(1-e^{-y-i\alpha}\right)  + \ln\left(1-e^{-y+i\alpha}\right)\right)\Bigg].
\end{align}
\eqref{ferdualm} may be regarded as the prediction of duality for the
off-shell free energy of the RB theory in the Higgsed phase.

\section{The tadpole from $W$ boson loops}\label{app}
The exact all-orders propagator $G_{\mu\nu}(q)$ in \eqref{exptcn} is
the saddle point value in the large $N_B$ limit of a gauge-singlet
field $\alpha_{\mu\nu}(q)$ that appears in \cite{Choudhury:2018iwf} as
one of two gauge-singlet fields $\alpha_{\mu\nu}$ and
$\Sigma^{\mu\nu}$ that describe the effective dynamics of the $W$
boson:
\begin{equation}
G_{\mu\nu}(q) = -\frac{1}{\lambda_B} \alpha_{\mu\nu}(q)\ .
\end{equation}
The first term of the gap equation \eqref{vgap} is then given by the
tadpole contribution
\begin{equation}\label{vgapalpha}
\lambda_B N_B  T(\sigma_B) =  \frac{\lambda_B N_B}{2\pi} \int \frac{\td^3 q}{(2\pi)^3} g^{\mu\rho} G(q)_{\rho\mu} = -\frac{N_B}{2\pi} \int \frac{\td^3 q}{(2\pi)^3} g^{\mu\rho} \alpha(q)_{\rho\mu}\ .
\end{equation}
For brevity, we work with the parameter $m = \lambda_B\sigma_B$ in
what follows. The field $\alpha_{\mu\nu}(q)$ is given in terms of four
known functions $F_1$, $F_2$, $F_3$ and $F_4$ of $w = 2q_+ q_-$ and
the all-loop exact kernel $Q(q) = G^{-1}(q)$.
\begin{align}\label{alphasol}
  \alpha_{++}(q) &= \frac{\lambda_B}{(2\pi)^2\detr Q}\, \frac{1}{q_-^2}\left(imF_1 + (F_3 + \tfrac{w}{2})^2\right)\ ,\nonumber\\
  \alpha_{-+}(q) &=  \frac{\lambda_B}{(2\pi)^2\detr Q}\left((1 - F_4)(F_3+\tfrac{w}{2}) - im( F_2 + im - q_3)\right) = \alpha_{+-}(-q)\ ,\nonumber\\
  \alpha_{--}(q) &=  \frac{\lambda_B}{(2\pi)^2\detr Q}\, q_-^2 (1 - F_4)^2\ ,\nonumber\\
  \alpha_{-3}(q) &=  -\frac{\lambda_B}{(2\pi)^2\detr Q}\, q_- (1 - F_4) (F_2 + i m - q_3) = \alpha_{3-}(-q)\ ,\nonumber\\
  \alpha_{3+}(q) &=  -\frac{\lambda_B}{(2\pi)^2 \detr Q} \frac{1}{q_-}\left( F_1 (1-F_4) + (F_2 + i m - q_3) (F_3 + \tfrac{w}{2})\right) = \alpha_{+3}(-q)\ ,\nonumber\\
  \alpha_{33}(q) &= -\frac{\lambda_B}{(2\pi)^2\detr Q} \left((F_2 + im)^2 - q_3^2\right)\ .
\end{align}
We give explicit expressions for the functions $F_{1\ldots 4}$:
\begin{align}\label{soln}
&F_2(w) = i m (g(w)-1)\ ,\quad F_4(w) = 1-{1\over g(w)}\ ,\nonumber\\
&F_3(w) = -\frac{w}{2} +\frac{1}{g(w)}\left(\frac{1}{2}\mathcal{I}(w)-\frac{m^2}{3} ( g(w)^3-g(0)^3 ) \right)\ ,\nonumber\\
&F_1(w) = im g(w) \left( c_B^2(g(w)-g(0))-\frac{m^2}{3}(g(w)^3-g(0)^3)+ w g(w)-\mathcal{I}(w)  \right)\ ,
\end{align}
where the functions $g(w)$ and $\mc{I}(w)$ are given by
\begin{equation}\label{gIdef}
  g(w) = 1 + \lambda_B \xi(w)\ ,\quad \mc{I}(w) = \int_0^w dz\, g(z) = w + \lambda_B \mc{I}_\xi(w)\ ,
\end{equation}
and the function $\xi(w)$ in the definition of $g(w)$ above is given
by
\begin{multline}\label{regintchi}
\xi(w) = \frac{1}{2 m \beta}\int_{-\pi}^\pi d\alpha\, \rho_B(\alpha) \bigg[\log 2\sinh\left(\tfrac{\beta}{2}\sqrt{w+c_B^2}+\tfrac{i}{2}\alpha\right) \\ + \log 2\sinh\left(\tfrac{\beta}{2}\sqrt{w+c_B^2}-\tfrac{i}{2}\alpha\right) \bigg]\ .
\end{multline}
We recognise the quantity $\cS$ defined in \eqref{nss} to be the value
of $\beta m\,\xi(w)$ at $w = 0$:
\begin{equation}
  \cS = \beta m\, \xi(0) = \frac{1}{2}\int_{-\pi}^\pi d\alpha\, \rho_B(\alpha) \bigg[\log 2\sinh\left(\tfrac{{\hat c}_B + i \alpha}{2}\right) + \log 2\sinh\left(\tfrac{{\hat c}_B - i\alpha}{2}\right) \bigg]\ .
\end{equation}
In the above expressions, the constant $c_B$ is the pole mass of the
$W$ boson which occurs in the determinant of the all-loop kernel
$Q(q) = G^{-1}(q)$:
\begin{equation}\label{detQ}
\det Q(q) = - \frac{m}{(2\pi)^3} (q^2 + c_B^2)\ ,
\end{equation}
and is given in terms of the parameter $m$ and the function $\xi(w)$
above by
\begin{equation}\label{Mexpr}
  c_B^2 =  m^2 ( 1 + \lambda_B \xi(0))^2\quad\text{or equivalently}\quad  \beta^2 c_B^2 = {\hat c}_B^2 =  ( {\hat m} + \lambda_B \cS)^2\ .
\end{equation}
Substituting the expressions \eqref{alphasol}, \eqref{soln} for
$\alpha_{\mu\nu}(q)$ in \eqref{vgapalpha}, we have
\begin{align}\label{detderv}
T(m) &= -\frac{1}{2\pi\lambda_B} \int \frac{\td^3q}{(2 \pi)^3} (\alpha_{+-}(q)  + \alpha_{-+}(q) + \alpha_{33}(q))\ ,\nonumber\\
&=\frac{1}{m} \int \frac{\td^3 q}{(2\pi)^3} \frac{1}{q^2 + c_B^2} \left(\frac{\mc{I}(w)}{g(w)^2}  + \frac{4 m^2}{3}g(w) + m^2 g(w)^2 + \frac{2m^2}{3} \frac{g(0)^3}{g(w)^2} + q_3^2\right)\ ,\nonumber\\
&=\frac{1}{m} \int \frac{\td^3 q}{(2\pi)^3} \frac{1}{q^2 + c_B^2} \left(\mc{L}(w)  - (w + c_B^2) \right) + \frac{1}{m} \int \frac{\td^3 q}{(2\pi)^3}\ ,
\end{align}
where we have added and subtracted the term $w + c_B^2$ inside the
integrand to complete the quantity $q_3^2$ to $q^2 + c_B^2$. The quantity
$\mc{L}(w)$ is given by
\begin{align} \label{Lform}
\mathcal{L}(w) &= {\mathcal{I}(w)\over g(w)^2}+{4\over 3} m^2 g(w)+ m^2 g(w)^2+{2\over 3}m^2{g(0)^3\over g(w)^2}\ .
\end{align}
The discrete sum over $q_3$ in the first term is given in terms of the function $\chi(w)$
\begin{align}\label{intdet}
\chi(w) & = -\frac{(2\pi)^3}{m}\int \frac{\td q_3}{2\pi} \frac{1}{q_3^2 + w + c_B^2} =  -\frac{(2\pi)^3}{m \beta} \int_{-\pi}^\pi d\alpha \rho_B(\alpha)\, \sum_{n\,\in\,\mathbb{Z}} \frac{1}{(2 \pi \frac{n}{\beta}+\frac{\alpha}{\beta})^2+w+c_B^2}\ ,\nonumber\\
&= -\frac{2\pi^3}{m} \int_{-\pi}^\pi d\alpha \rho_B(\alpha) \frac{1}{\sqrt{w + c_B^2}}\times \nonumber \\ &\qquad\qquad\qquad \times \left(\coth \left(\tfrac{\beta}{2}\sqrt{w + c_B^2} + \tfrac{i}{2}\alpha\right) + \coth \left(\tfrac{\beta}{2}\sqrt{w + c_B^2} - \tfrac{i}{2}\alpha\right)\right)\ . 
\end{align}
The $q_3$ sum in the last term in \eqref{detderv} is given by
$c_0 \equiv \textstyle \sum_n 1$ and is hence divergent. We regularise
the divergent sum using $\zeta$-function regularisation in which case
we have $c_0(\text{reg}.) = 1 + 2\zeta(0) = 0$. Thus, equation \eqref{detderv} becomes
\begin{align}\label{detderv2}
 T(m) &= -\frac{1}{(2\pi)^3}\int_0^{\infty} \frac{dw}{4\pi} \ \chi(w) \left(\mathcal{L}(w) -(w+c_B^2)\right) = 2\int_{0}^{\infty} \frac{dw}{4\pi}\, \xi'(w) (\mathcal{L}(w) - (w+c_B^2))\ ,
\end{align}
where we have used $\chi(w)=-2(2\pi)^3\xi'(w)$. Next, recall the
expressions \eqref{gIdef} and \eqref{Mexpr}:
\begin{equation} \label{pte}
g(w) = 1 + \lambda_B \xi(w)\ ,\quad \mc{I}(w) = \int_0^w dz g(z) = w + \lambda_B \mc{I}_\xi(w)\ ,\quad c_B^2 = m^2 (1 + \lambda_B \xi(0))^2\ .
\end{equation}
Inserting \eqref{pte} into \eqref{detderv2} and Taylor-expanding $\mc{L}(w)-(w+c_B^2)$ in all explicit factors of $\lambda_B$ (around $\lambda_B = 0$) we find 
\begin{align}
\mc{L}(w) -(w + c_B^2)  &= {\mathcal{I}(w)\over g(w)^2}+{4\over 3} m^2 g(w)+ m^2 g(w)^2+{2\over 3}m^2{g(0)^3\over g(w)^2} - (w + m^2 g(0)^2)\ ,\nonumber\\
           &= \sum_{n= 0}^\infty (-\lambda_B)^n \mc{L}_n(w)\ ,
\end{align}
with
\begin{align}\label{Lnexpn}
 &\mathcal{L}_0(w)= 2m^2\ ,\quad \mathcal{L}_1(w)= -\left(\mathcal{I}_\xi(w) -2 w \xi (w)\right) - 2m^2 \xi (w)\ ,\nonumber\\
  &\mathcal{L}_2(w)= \left(-2 \mathcal{I}_\xi(w) \xi (w)+3 w \xi (w)^2\right) + m^2 \left(3\xi(w)^2-4\xi(w)\xi(0)+\xi(0)^2\right)\ ,\nonumber\\
&\mathcal{L}_n(w)= \left((n+1) w \xi(w)^n - n \xi(w)^{n-1} \mathcal{I}_\xi(w)\right) + \nonumber\\ &\frac{2 m^2}{3}\Big((n+1)\xi(w)^n  - (n-2)\xi(w)^{n-3}\xi(0)^3 + 3(n-1)\xi(w)^{n-2}\xi(0)^2-3n \ \xi(w)^{n-1}\xi(0)\Big) \quad\text{for}\quad n\geq 3\ .
\end{align}
The integral over $w$ in \eqref{detderv2} becomes
\begin{equation}
\sum_{n=0}^\infty (-\lambda_B)^n \int_0^\infty \frac{dw}{4\pi} \xi'(w) \mc{L}_n(w)\ .
\end{equation}

The integral over the first two terms in the expressions for
${\cal L}_{n\geq 1}$ in \eqref{Lnexpn} can be simplified by writing
this in a total derivative form
\begin{equation}
\begin{aligned}
dw \ \xi '(w) \left(( n+1)  \xi (w)^n w-n  \xi (w)^{n-1} \mathcal{I}_\xi(w)\right)=d( \xi (w)^{n+1} w-\xi (w)^{n} \mathcal{I}_\xi(w))
\end{aligned}
\end{equation}
In the dimensional regularisation scheme used in our previous paper
\cite{Choudhury:2018iwf} we have $\xi(\infty )=0$. Also, by definition
we have $\mathcal{I}_\xi(0)=0$. This implies that
\begin{equation}
\begin{aligned}
\int_0^\infty dw \ \xi '(w) \left(( n+1)  \xi (w)^n w-n  \xi (w)^{n-1} \mathcal{I}_\xi(w)\right)=0\ .
\end{aligned}
\end{equation}
The remaining terms in ${\cal L}$ are simple polynomials in $\xi$ and
the integrations can be easily performed. Only $\mc{L}_0$ and $\mc{L}_1$ give non-zero contributions:
\begin{align} \label{Lns}
  &\int_0^\infty dw  \ \xi'(w) \mathcal{L}_0(w)= -2 m^2 \xi(0)\ ,\quad \int_0^\infty dw  \ \xi'(w) \mathcal{L}_1(w)= + m^2 \xi(0)^2\ .
\end{align}
Substituting the above results into \eqref{detderv2}, we get
\begin{equation}
T(m) = -\frac{m^2}{2\pi} \Big(2\xi(0)+\lambda_B \xi(0)^2\Big)\ .
\end{equation}
Recalling the equation \eqref{Mexpr} for ${\hat c}_B$ and
$m = \lambda_B \sigma_B$, we have the final expression for the tadpole
contribution from the $W$ boson propagator \eqref{vgapalpha}:
\begin{equation}\label{detdervfinal}
\lambda_B N_B T(\sigma_B) = -\frac{N_B}{2\pi} \left({ c}_B^2-\lambda_B^2 {\sigma}_B^2 \right)\ .
\end{equation}

\section{The critical boson scaling limit} \label{cbl}
Recall that the RB theory reduces to the critical boson or 
CB theory in the scaling limit 
\begin{equation}\label{cbsl}
m_B^2 \to \infty, ~~~\lambda_B b_4 \to \infty, ~~~\frac{m_B^2}{2 \lambda_B b_4}= m_{B}^{\rm cri}= {\rm fixed}\ .
\end{equation}
In this subsection we study the reduction of the off-shell free
energy under this scaling limit. The off-shell free energy
\eqref{Feffofsh} simplifies in the limit \eqref{cbsl} as follows. The
second term in the second line of \eqref{Feffofsh} reduces to
\begin{equation}
\label{eesl}
6 \lambda_B {\hat b}_4 {\hat \sigma}_B 
\left({\hat m}_B^{\rm cri} + {\hat \sigma}_B \right)
= 6 \lambda_B {\hat b}_4 \left[ 
\left({\hat \sigma}_B +  \frac{{\hat m}_B^{\rm cri}}{2} \right)^2-
\frac{({\hat m}_B^{\rm cri})^2}{4} \right]
\end{equation}
Note that confining potential \eqref{eesl} is infinitely stiff in the
CB scaling limit. It follows that ${\hat \sigma}_B$ is frozen at the
minimum of \eqref{eesl} i.e. at ${\sigma}_B= -\frac{m_B^{\rm cri}}{2}$
in the CB scaling limit. It follows that in this limit
\eqref{Feffofsh} simplifies to
\begin{align}\label{Fsl}
F(c_B, \tl\cS) & = \frac{N_B }{6\pi} \Bigg[- {\hat c}_B^3- 4  {\tilde\cS}^3 \lambda_B^2 + \frac{3}{2} {\hat c}_B^2 {\hat m}_B^{\rm cri} \nonumber\\
&\qquad\qquad+ 6{\tilde\cS}^2 \lambda_B^2 {\hat m}_B^{\rm cri} -3{\tilde\cS} \lambda_B^2 ({\hat m}_B^{\rm cri})^2  +6 {\hat c}_B |\lambda_B| ({\tilde\cS} - 
\tfrac{{\hat m}_B^{\rm cri}}{2})^2  \nonumber\\
&\qquad\qquad + 3 \int_{-\pi}^{\pi}d\alpha \rho_{B}(\alpha)\int_{{\hat c}_B}^{\infty} dy ~y~\left( \ln\left(1-e^{-y-i\alpha}\right)  + \ln\left(1-e^{-y+i\alpha}\right)\right)\Bigg].
\end{align}
(we have omitted a divergent constant proportional to $b_4$ that can
be cancelled by a cosmological constant counterterm.) Extremizing
\eqref{Fsl} w.r.t. ${\tilde \cS}$ we recover the first of the gap
equations in \eqref{foll} under the replacement
$${\sigma}_B \rightarrow-\frac{m_B^{\rm cri}}{2}\ .$$  The two inequivalent
solutions of this equation are \eqref{foee} and \eqref{soee} under the
same replacement for ${\sigma}_B$. These solutions correspond to the
unHiggsed and Higgsed branches respectively.

On the Higgsed branch we can plug the solution of \eqref{soee} back
into \eqref{Fsl} to find a free energy as a function of the single
off-shell variable $c_B$; the final result of this exercise is given
by the critical boson free energy given in
\eqref{osfeeh}\footnote{This procedure automatically produces a
	particular choice of the cosmological constant counterterm. It would
	be interesting to investigate if this particular value has physical
	significance.}. In a similar manner, on the unHiggsed branch we can
plug the solution of \eqref{foee} into \eqref{Fsl} to find off-shell
free energy as a function of $c_B$, given by
\begin{equation} \begin{split}
\label{oscs}
F_{\rm CB}(c_B) &=\frac{N_B}{6\pi} {\Bigg[}  -
{\hat c}_B^3 +\frac{3}{2} {\hat m}_B^{\rm cri} {\hat c}_B^2
-\frac{\lambda_B^2 ({\hat m}_B^{\rm cri})^3}{2} \\
&\qquad\quad+3 \int_{-\pi}^{\pi} \rho(\alpha) d\alpha\int_{{\hat c}_B}^{\infty}dy y\left(\ln\left(1-e^{-y-i\alpha}\right)+\ln\left(1-e^{-y+i\alpha}\right)  \right)
{\Bigg]}\ .
\end{split}
\end{equation}

\bigskip

\providecommand{\href}[2]{#2}\begingroup\raggedright\endgroup


\begin{thebibliography}{10}

\bibitem{Choudhury:2018iwf}
S.~Choudhury, A.~Dey, I.~Halder, S.~Jain, L.~Janagal, S.~Minwalla, and
  N.~Prabhakar, {\it {Bose-Fermi Chern-Simons Dualities in the Higgsed Phase}},
   [\href{http://xxx.lanl.gov/abs/1804.08635}{{\tt arXiv:1804.08635}}].

\bibitem{Minwalla:2015sca}
S.~Minwalla and S.~Yokoyama, {\it {Chern Simons Bosonization along RG Flows}},
  {\em JHEP} {\bf 02} (2016) 103,
  [\href{http://xxx.lanl.gov/abs/1507.04546}{{\tt arXiv:1507.04546}}].

\bibitem{abcd} O.~Aharony, S.~Jain, and S.~Minwalla, {\it Flows, Fixed
    Points and Duality in Chern-Simons-matter theories},
  [\href{http://arxiv.org/abs/1808.03317}{{\tt arXiv:1808.03317}}].

\bibitem{Aharony:2012nh}
O.~Aharony, G.~Gur-Ari, and R.~Yacoby, {\it {Correlation Functions of Large N
  Chern-Simons-Matter Theories and Bosonization in Three Dimensions}},  {\em
  JHEP} {\bf 1212} (2012) 028, [\href{http://xxx.lanl.gov/abs/1207.4593}{{\tt
  arXiv:1207.4593}}].

\bibitem{Jain:2014nza}
S.~Jain, M.~Mandlik, S.~Minwalla, T.~Takimi, S.~R. Wadia, and S.~Yokoyama, {\it
  {Unitarity, Crossing Symmetry and Duality of the S-matrix in large N
  Chern-Simons theories with fundamental matter}},  {\em JHEP} {\bf 04} (2015)
  129, [\href{http://xxx.lanl.gov/abs/1404.6373}{{\tt arXiv:1404.6373}}].

\bibitem{Yokoyama:2016sbx}
S.~Yokoyama, {\it {Scattering Amplitude and Bosonization Duality in General
  Chern-Simons Vector Models}},  {\em JHEP} {\bf 09} (2016) 105,
  [\href{http://xxx.lanl.gov/abs/1604.01897}{{\tt arXiv:1604.01897}}].

\bibitem{Jain:2013py}
S.~Jain, S.~Minwalla, T.~Sharma, T.~Takimi, S.~R. Wadia, et~al., {\it {Phases
  of large $N$ vector Chern-Simons theories on $S^2 x S^1$}},  {\em JHEP} {\bf
  1309} (2013) 009, [\href{http://xxx.lanl.gov/abs/1301.6169}{{\tt
  arXiv:1301.6169}}].

\bibitem{Giombi:2017rhm}
S.~Giombi, V.~Kirilin, and E.~Skvortsov, {\it {Notes on Spinning Operators in
  Fermionic CFT}},  {\em JHEP} {\bf 05} (2017) 041,
  [\href{http://xxx.lanl.gov/abs/1701.06997}{{\tt arXiv:1701.06997}}].

\bibitem{Komargodski:2017keh}
Z.~Komargodski and N.~Seiberg, {\it {A symmetry breaking scenario for
  QCD$_{3}$}},  {\em JHEP} {\bf 01} (2018) 109,
  [\href{http://xxx.lanl.gov/abs/1706.08755}{{\tt arXiv:1706.08755}}].

\bibitem{Sezgin:2017jgm}
E.~Sezgin, E.~D. Skvortsov, and Y.~Zhu, {\it {Chern-Simons Matter Theories and
  Higher Spin Gravity}},  {\em JHEP} {\bf 07} (2017) 133,
  [\href{http://xxx.lanl.gov/abs/1705.03197}{{\tt arXiv:1705.03197}}].

\bibitem{Aitken:2017nfd}
K.~Aitken, A.~Baumgartner, A.~Karch, and B.~Robinson, {\it {3d Abelian
  Dualities with Boundaries}},  {\em JHEP} {\bf 03} (2018) 053,
  [\href{http://xxx.lanl.gov/abs/1712.02801}{{\tt arXiv:1712.02801}}].

\bibitem{Karch:2018mer}
A.~Karch, D.~Tong, and C.~Turner, {\it {Mirror Symmetry and Bosonization in 2d
  and 3d}},  {\em JHEP} {\bf 07} (2018) 059,
  [\href{http://xxx.lanl.gov/abs/1805.00941}{{\tt arXiv:1805.00941}}].

\bibitem{Aharony:2018npf}
O.~Aharony, L.~F. Alday, A.~Bissi, and R.~Yacoby, {\it {The Analytic Bootstrap
  for Large $N$ Chern-Simons Vector Models}},
  [\href{http://xxx.lanl.gov/abs/1805.04377}{{\tt arXiv:1805.04377}}].

\bibitem{Yacoby:2018yvy}
R.~Yacoby, {\it {Scalar Correlators in Bosonic Chern-Simons Vector Models}},
  [\href{http://xxx.lanl.gov/abs/1805.11627}{{\tt arXiv:1805.11627}}].

\bibitem{Aitken:2018cvh}
K.~Aitken, A.~Baumgartner, and A.~Karch, {\it {Novel 3d bosonic dualities from
  bosonization and holography}},
  [\href{http://xxx.lanl.gov/abs/1807.01321}{{\tt arXiv:1807.01321}}].

\bibitem{Aharony:2003sx}
O.~Aharony, J.~Marsano, S.~Minwalla, K.~Papadodimas, and M.~Van~Raamsdonk, {\it
  {The Hagedorn - deconfinement phase transition in weakly coupled large N
  gauge theories}},  {\em Adv. Theor. Math. Phys.} {\bf 8} (2004) 603--696,
  [\href{http://xxx.lanl.gov/abs/hep-th/0310285}{{\tt hep-th/0310285}}].
  [,161(2003)].

\bibitem{Klebanov:2002ja}
I.~R. Klebanov and A.~M. Polyakov, {\it {AdS dual of the critical O(N) vector
  model}},  {\em Phys. Lett.} {\bf B550} (2002) 213--219,
  [\href{http://xxx.lanl.gov/abs/hep-th/0210114}{{\tt hep-th/0210114}}].

\bibitem{Sezgin:2002rt}
E.~Sezgin and P.~Sundell, {\it {Massless higher spins and holography}},  {\em
  Nucl. Phys.} {\bf B644} (2002) 303--370,
  [\href{http://xxx.lanl.gov/abs/hep-th/0205131}{{\tt hep-th/0205131}}].
  [Erratum: Nucl. Phys.B660,403(2003)].

\bibitem{Giombi:2009wh}
S.~Giombi and X.~Yin, {\it {Higher Spin Gauge Theory and Holography: The
  Three-Point Functions}},  {\em JHEP} {\bf 09} (2010) 115,
  [\href{http://xxx.lanl.gov/abs/0912.3462}{{\tt arXiv:0912.3462}}].

\bibitem{Giombi:2011kc}
S.~Giombi, S.~Minwalla, S.~Prakash, S.~P. Trivedi, S.~R. Wadia, and X.~Yin,
  {\it {Chern-Simons Theory with Vector Fermion Matter}},  {\em Eur. Phys. J.}
  {\bf C72} (2012) 2112, [\href{http://xxx.lanl.gov/abs/1110.4386}{{\tt
  arXiv:1110.4386}}].

\bibitem{Chang:2012kt}
C.-M. Chang, S.~Minwalla, T.~Sharma, and X.~Yin, {\it {ABJ Triality: from
  Higher Spin Fields to Strings}},  {\em J. Phys.} {\bf A46} (2013) 214009,
  [\href{http://xxx.lanl.gov/abs/1207.4485}{{\tt arXiv:1207.4485}}].

\bibitem{Maldacena:2011jn}
J.~Maldacena and A.~Zhiboedov, {\it {Constraining Conformal Field Theories with
  A Higher Spin Symmetry}},  {\em J.Phys.} {\bf A46} (2013) 214011,
  [\href{http://xxx.lanl.gov/abs/1112.1016}{{\tt arXiv:1112.1016}}].

\bibitem{Maldacena:2012sf}
J.~Maldacena and A.~Zhiboedov, {\it {Constraining conformal field theories with
  a slightly broken higher spin symmetry}},  {\em Class.Quant.Grav.} {\bf 30}
  (2013) 104003, [\href{http://xxx.lanl.gov/abs/1204.3882}{{\tt
  arXiv:1204.3882}}].

\bibitem{GurAri:2012is}
G.~Gur-Ari and R.~Yacoby, {\it {Correlators of Large N Fermionic Chern-Simons
  Vector Models}},  {\em JHEP} {\bf 1302} (2013) 150,
  [\href{http://xxx.lanl.gov/abs/1211.1866}{{\tt arXiv:1211.1866}}].

\bibitem{Bedhotiya:2015uga}
A.~Bedhotiya and S.~Prakash, {\it {A test of bosonization at the level of
  four-point functions in Chern-Simons vector models}},  {\em JHEP} {\bf 12}
  (2015) 032, [\href{http://xxx.lanl.gov/abs/1506.05412}{{\tt
  arXiv:1506.05412}}].

\bibitem{Turiaci:2018nua}
G.~J. Turiaci and A.~Zhiboedov, {\it {Veneziano Amplitude of Vasiliev Theory}},
   [\href{http://xxx.lanl.gov/abs/1802.04390}{{\tt arXiv:1802.04390}}].

\bibitem{Seiberg:2016gmd}
N.~Seiberg, T.~Senthil, C.~Wang, and E.~Witten, {\it {A Duality Web in 2+1
  Dimensions and Condensed Matter Physics}},
  [\href{http://xxx.lanl.gov/abs/1606.01989}{{\tt arXiv:1606.01989}}].

\bibitem{Murugan:2016zal}
J.~Murugan and H.~Nastase, {\it {Particle-vortex duality in topological
  insulators and superconductors}},  {\em JHEP} {\bf 05} (2017) 159,
  [\href{http://xxx.lanl.gov/abs/1606.01912}{{\tt arXiv:1606.01912}}].

\bibitem{Yokoyama:2012fa}
S.~Yokoyama, {\it {Chern-Simons-Fermion Vector Model with Chemical Potential}},
   {\em JHEP} {\bf 1301} (2013) 052,
  [\href{http://xxx.lanl.gov/abs/1210.4109}{{\tt arXiv:1210.4109}}].

\bibitem{Jain:2012qi}
S.~Jain, S.~P. Trivedi, S.~R. Wadia, and S.~Yokoyama, {\it {Supersymmetric
  Chern-Simons Theories with Vector Matter}},  {\em JHEP} {\bf 1210} (2012)
  194, [\href{http://xxx.lanl.gov/abs/1207.4750}{{\tt arXiv:1207.4750}}].

\bibitem{Aharony:2012ns}
O.~Aharony, S.~Giombi, G.~Gur-Ari, J.~Maldacena, and R.~Yacoby, {\it {The
  Thermal Free Energy in Large N Chern-Simons-Matter Theories}},  {\em JHEP}
  {\bf 1303} (2013) 121, [\href{http://xxx.lanl.gov/abs/1211.4843}{{\tt
  arXiv:1211.4843}}].

\bibitem{Takimi:2013zca}
T.~Takimi, {\it {Duality and higher temperature phases of large N Chern-Simons
  matter theories on $S^2$ x $S^1$}},  {\em JHEP} {\bf 1307} (2013) 177,
  [\href{http://xxx.lanl.gov/abs/1304.3725}{{\tt arXiv:1304.3725}}].

\bibitem{Yokoyama:2013pxa}
S.~Yokoyama, {\it {A Note on Large N Thermal Free Energy in Supersymmetric
  Chern-Simons Vector Models}},  {\em JHEP} {\bf 1401} (2014) 148,
  [\href{http://xxx.lanl.gov/abs/1310.0902}{{\tt arXiv:1310.0902}}].

\bibitem{Jain:2013gza}
S.~Jain, S.~Minwalla, and S.~Yokoyama, {\it {Chern Simons duality with a
  fundamental boson and fermion}},  {\em JHEP} {\bf 1311} (2013) 037,
  [\href{http://xxx.lanl.gov/abs/1305.7235}{{\tt arXiv:1305.7235}}].

\bibitem{xyz}
G.~GurAri and R.~Yacoby, {\it {Three Dimensional Bosonization From
  Supersymmetry}},  {\em JHEP} {\bf 11} (2015) 013,
  [\href{http://xxx.lanl.gov/abs/1507.04378}{{\tt arXiv:1507.04378}}].

\bibitem{Geracie:2015drf}
M.~Geracie, M.~Goykhman, and D.~T. Son, {\it {Dense Chern-Simons Matter with
  Fermions at Large N}},  {\em JHEP} {\bf 04} (2016) 103,
  [\href{http://xxx.lanl.gov/abs/1511.04772}{{\tt arXiv:1511.04772}}].

\bibitem{Inbasekar:2015tsa}
K.~Inbasekar, S.~Jain, S.~Mazumdar, S.~Minwalla, V.~Umesh, and S.~Yokoyama,
  {\it {Unitarity, crossing symmetry and duality in the scattering of $
  \mathcal{N}=1 $ susy matter Chern-Simons theories}},  {\em JHEP} {\bf 10}
  (2015) 176, [\href{http://xxx.lanl.gov/abs/1505.06571}{{\tt
  arXiv:1505.06571}}].

\bibitem{Dandekar:2014era}
Y.~Dandekar, M.~Mandlik, and S.~Minwalla, {\it {Poles in the $S$-Matrix of
  Relativistic Chern-Simons Matter theories from Quantum Mechanics}},  {\em
  JHEP} {\bf 04} (2015) 102, [\href{http://xxx.lanl.gov/abs/1407.1322}{{\tt
  arXiv:1407.1322}}].

\bibitem{Inbasekar:2017ieo}
K.~Inbasekar, S.~Jain, P.~Nayak, and V.~Umesh, {\it {All tree level scattering
  amplitudes in Chern-Simons theories with fundamental matter}},
  [\href{http://xxx.lanl.gov/abs/1710.04227}{{\tt arXiv:1710.04227}}].

\bibitem{Inbasekar:2017sqp}
K.~Inbasekar, S.~Jain, S.~Majumdar, P.~Nayak, T.~Neogi, T.~Sharma, R.~Sinha,
  and V.~Umesh, {\it {Dual Superconformal Symmetry of ${\cal N}=2$ Chern-Simons
  theory with Fundamental Matter and Non-Renormalization at Large $N$}},
  [\href{http://xxx.lanl.gov/abs/1711.02672}{{\tt arXiv:1711.02672}}].

\bibitem{Benini:2011mf}
F.~Benini, C.~Closset, and S.~Cremonesi, {\it {Comments on 3d Seiberg-like
  dualities}},  {\em JHEP} {\bf 1110} (2011) 075,
  [\href{http://xxx.lanl.gov/abs/1108.5373}{{\tt arXiv:1108.5373}}].

\bibitem{Gur-Ari:2016xff}
G.~Gur-Ari, S.~A. Hartnoll, and R.~Mahajan, {\it {Transport in
  Chern-Simons-Matter Theories}},  {\em JHEP} {\bf 07} (2016) 090,
  [\href{http://xxx.lanl.gov/abs/1605.01122}{{\tt arXiv:1605.01122}}].

\bibitem{Aharony:2011jz}
O.~Aharony, G.~GurAri, and R.~Yacoby, {\it {d=3 Bosonic Vector Models Coupled
  to Chern-Simons Gauge Theories}},  {\em JHEP} {\bf 1203} (2012) 037,
  [\href{http://xxx.lanl.gov/abs/1110.4382}{{\tt arXiv:1110.4382}}].

\bibitem{Nosaka:2017ohr}
T.~Nosaka and S.~Yokoyama, {\it {Complete factorization in minimal N=4
  Chern-Simons-matter theory}},
  [\href{http://xxx.lanl.gov/abs/1706.07234}{{\tt arXiv:1706.07234}}].

\bibitem{Giombi:2016ejx}
S.~Giombi, {\it {Higher Spin - CFT Duality}},  in {\em {Proceedings,
  Theoretical Advanced Study Institute in Elementary Particle Physics: New
  Frontiers in Fields and Strings (TASI 2015): Boulder, CO, USA, June 1-26,
  2015}}, pp.~137--214, 2017.
\newblock \href{http://xxx.lanl.gov/abs/1607.02967}{{\tt arXiv:1607.02967}}.

\bibitem{Wadia:2016zpd}
S.~R. Wadia, {\it {Chern-Simons theories with fundamental matter: A brief
  review of large $N$ results including Fermi-Bose duality and the S-matrix}},
  {\em Int. J. Mod. Phys.} {\bf A31} (2016), no.~32 1630052.

\bibitem{Frishman:2014cma}
Y.~Frishman and J.~Sonnenschein, {\it {Large N Chern-Simons with massive
  fundamental fermions - A model with no bound states}},  {\em JHEP} {\bf 1412}
  (2014) 165, [\href{http://xxx.lanl.gov/abs/1409.6083}{{\tt
  arXiv:1409.6083}}].

\bibitem{Gurucharan:2014cva}
V.~Gurucharan and S.~Prakash, {\it {Anomalous dimensions in non-supersymmetric
  bifundamental Chern-Simons theories}},  {\em JHEP} {\bf 1409} (2014) 009,
  [\href{http://xxx.lanl.gov/abs/1404.7849}{{\tt arXiv:1404.7849}}].

\bibitem{Giombi:2017txg}
S.~Giombi, {\it {Testing the Boson/Fermion Duality on the Three-Sphere}},
  [\href{http://xxx.lanl.gov/abs/1707.06604}{{\tt arXiv:1707.06604}}].

\bibitem{Bardeen:2014qua}
W.~A. Bardeen, {\it {The Massive Fermion Phase for the U(N) Chern-Simons Gauge
  Theory in D=3 at Large N}},  {\em JHEP} {\bf 1410} (2014) 39,
  [\href{http://xxx.lanl.gov/abs/1404.7477}{{\tt arXiv:1404.7477}}].

\bibitem{Bardeen:2014paa}
W.~A. Bardeen and M.~Moshe, {\it {Spontaneous breaking of scale invariance in a
  D=3 U(N ) model with Chern-Simons gauge fields}},  {\em JHEP} {\bf 1406}
  (2014) 113, [\href{http://xxx.lanl.gov/abs/1402.4196}{{\tt
  arXiv:1402.4196}}].

\bibitem{Moshe:2014bja}
M.~Moshe and J.~Zinn-Justin, {\it {3D Field Theories with Chern--Simons Term
  for Large $N$ in the Weyl Gauge}},  {\em JHEP} {\bf 1501} (2015) 054,
  [\href{http://xxx.lanl.gov/abs/1410.0558}{{\tt arXiv:1410.0558}}].

\bibitem{Giombi:2016zwa}
S.~Giombi, V.~Gurucharan, V.~Kirilin, S.~Prakash, and E.~Skvortsov, {\it {On
  the Higher-Spin Spectrum in Large N Chern-Simons Vector Models}},  {\em JHEP}
  {\bf 01} (2017) 058, [\href{http://xxx.lanl.gov/abs/1610.08472}{{\tt
  arXiv:1610.08472}}].

\bibitem{Charan:2017jyc}
V.~G. Charan and S.~Prakash, {\it {On the Higher Spin Spectrum of Chern-Simons
  Theory coupled to Fermions in the Large Flavour Limit}},  {\em JHEP} {\bf 02}
  (2018) 094, [\href{http://xxx.lanl.gov/abs/1711.11300}{{\tt
  arXiv:1711.11300}}].

\bibitem{Radicevic:2015yla}
D.~Radicevic, {\it {Disorder Operators in Chern-Simons-Fermion Theories}},
  {\em JHEP} {\bf 03} (2016) 131,
  [\href{http://xxx.lanl.gov/abs/1511.01902}{{\tt arXiv:1511.01902}}].

\bibitem{Aharony:2015mjs}
O.~Aharony, {\it {Baryons, monopoles and dualities in Chern-Simons-matter
  theories}},  {\em JHEP} {\bf 02} (2016) 093,
  [\href{http://xxx.lanl.gov/abs/1512.00161}{{\tt arXiv:1512.00161}}].

\bibitem{Aharony:2015pla}
O.~Aharony, P.~Narayan, and T.~Sharma, {\it {On monopole operators in
  supersymmetric Chern-Simons-matter theories}},  {\em JHEP} {\bf 05} (2015)
  117, [\href{http://xxx.lanl.gov/abs/1502.00945}{{\tt arXiv:1502.00945}}].

\bibitem{Radicevic:2016wqn}
D.~Radicevic, D.~Tong, and C.~Turner, {\it {Non-Abelian 3d Bosonization and
  Quantum Hall States}},  {\em JHEP} {\bf 12} (2016) 067,
  [\href{http://xxx.lanl.gov/abs/1608.04732}{{\tt arXiv:1608.04732}}].

\bibitem{Aharony:2016jvv}
O.~Aharony, F.~Benini, P.-S. Hsin, and N.~Seiberg, {\it {Chern-Simons-matter
  dualities with $SO$ and $USp$ gauge groups}},  {\em JHEP} {\bf 02} (2017)
  072, [\href{http://xxx.lanl.gov/abs/1611.07874}{{\tt arXiv:1611.07874}}].

\bibitem{Karch:2016sxi}
A.~Karch and D.~Tong, {\it {Particle-Vortex Duality from 3d Bosonization}},
  {\em Phys. Rev.} {\bf X6} (2016), no.~3 031043,
  [\href{http://xxx.lanl.gov/abs/1606.01893}{{\tt arXiv:1606.01893}}].

\bibitem{Karch:2016aux}
A.~Karch, B.~Robinson, and D.~Tong, {\it {More Abelian Dualities in 2+1
  Dimensions}},  {\em JHEP} {\bf 01} (2017) 017,
  [\href{http://xxx.lanl.gov/abs/1609.04012}{{\tt arXiv:1609.04012}}].

\bibitem{Hsin:2016blu}
P.-S. Hsin and N.~Seiberg, {\it {Level/rank Duality and Chern-Simons-Matter
  Theories}},  {\em JHEP} {\bf 09} (2016) 095,
  [\href{http://xxx.lanl.gov/abs/1607.07457}{{\tt arXiv:1607.07457}}].

\bibitem{Gomis:2017ixy}
J.~Gomis, Z.~Komargodski, and N.~Seiberg, {\it {Phases Of Adjoint QCD$_3$ And
  Dualities}},  [\href{http://xxx.lanl.gov/abs/1710.03258}{{\tt
  arXiv:1710.03258}}].

\bibitem{Jensen:2017bjo}
K.~Jensen, {\it {A master bosonization duality}},  {\em JHEP} {\bf 01} (2018)
  031, [\href{http://xxx.lanl.gov/abs/1712.04933}{{\tt arXiv:1712.04933}}].

\bibitem{Gaiotto:2017tne}
D.~Gaiotto, Z.~Komargodski, and N.~Seiberg, {\it {Time-Reversal Breaking in
  QCD$_4$, Walls, and Dualities in 2+1 Dimensions}},
  [\href{http://xxx.lanl.gov/abs/1708.06806}{{\tt arXiv:1708.06806}}].

\bibitem{Benini:2017dus}
F.~Benini, P.-S. Hsin, and N.~Seiberg, {\it {Comments on global symmetries,
  anomalies, and duality in (2 + 1)d}},  {\em JHEP} {\bf 04} (2017) 135,
  [\href{http://xxx.lanl.gov/abs/1702.07035}{{\tt arXiv:1702.07035}}].

\bibitem{Jensen:2017xbs}
K.~Jensen and A.~Karch, {\it {Embedding three-dimensional bosonization
  dualities into string theory}},
  [\href{http://xxx.lanl.gov/abs/1709.07872}{{\tt arXiv:1709.07872}}].

\bibitem{Jensen:2017dso}
K.~Jensen and A.~Karch, {\it {Bosonizing three-dimensional quiver gauge
  theories}},  [\href{http://xxx.lanl.gov/abs/1709.01083}{{\tt
  arXiv:1709.01083}}].

\bibitem{Benini:2017aed}
F.~Benini, {\it {Three-dimensional dualities with bosons and fermions}},  {\em
  JHEP} {\bf 02} (2018) 068, [\href{http://xxx.lanl.gov/abs/1712.00020}{{\tt
  arXiv:1712.00020}}].

\bibitem{Cordova:2017vab}
C.~Cordova, P.-S. Hsin, and N.~Seiberg, {\it {Global Symmetries, Counterterms,
  and Duality in Chern-Simons Matter Theories with Orthogonal Gauge Groups}},
  [\href{http://xxx.lanl.gov/abs/1711.10008}{{\tt arXiv:1711.10008}}].

\end{thebibliography}
\end{document}